\documentclass[journal]{IEEEtran}
\usepackage{float}
\usepackage{hyperref}
\usepackage{comment}
\usepackage{gensymb} % for \degree
\usepackage[flushleft]{threeparttable}
\usepackage{multirow}
\usepackage{amsmath}
\usepackage{amssymb}
\usepackage{bm}
\usepackage{graphicx}
\usepackage{subfig}
\usepackage[linesnumbered,ruled,vlined]{algorithm2e}
\usepackage{algorithmic}
\usepackage[utf8]{inputenc}
\usepackage[T1]{fontenc}
\usepackage{textcomp}
\usepackage{nomencl}
\usepackage{rotating}
\usepackage{array}
\usepackage{mathtools}
\usepackage{diagbox}
\usepackage{url}
\usepackage{footnote}
\usepackage{slashbox}

\makesavenoteenv{tabular}
\makesavenoteenv{table}

\DeclareMathOperator\Arg{Arg}
\DeclareMathOperator{\arctantwo}{arctan2}

\makeatletter
\def\BState{\State\hskip-\ALG@thistlm}
\DeclarePairedDelimiter{\norm}{\lVert}{\rVert} 
\makeatother

\graphicspath{{figures/}}

\makenomenclature
\ifCLASSINFOpdf
\else
\fi
\begin{document}

%
% paper title
% Titles are generally capitalized except for words such as a, an, and, as,
% at, but, by, for, in, nor, of, on, or, the, to and up, which are usually
% not capitalized unless they are the first or last word of the title.
% Linebreaks \\ can be used within to get better formatting as desired.
% Do not put math or special symbols in the title.

\title{Multirotors From Takeoff to Real-Time Full Identification Using the Modified Relay Feedback Test and Deep Neural Networks}

%
%
% author names and IEEE memberships
% note positions of commas and nonbreaking spaces ( ~ ) LaTeX will not break
% a structure at a ~ so this keeps an author's name from being broken across
% two lines.
% use \thanks{} to gain access to the first footnote area
% a separate \thanks must be used for each paragraph as LaTeX2e's \thanks
% was not built to handle multiple paragraphs
%

\author{
        Abdulla~Ayyad\href{https://orcid.org/0000-0002-3006-2320}{\includegraphics[scale=0.75]{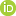}},~\IEEEmembership{Member,~IEEE,}
        Mohamad~Chehadeh\href{https://orcid.org/0000-0002-9430-3349}{\includegraphics[scale=0.75]{orcid.png}},~\IEEEmembership{Member,~IEEE,}
        Pedro~Silva\href{https://orcid.org/0000-0002-2598-7695}{\includegraphics[scale=0.75]{orcid.png}},
        Mohamad~Wahbah\href{https://orcid.org/0000-0003-1647-8546}{\includegraphics[scale=0.75]{orcid.png}},
        Oussama~Abdul~Hay\href{https://orcid.org/0000-0001-8299-6021}{\includegraphics[scale=0.75]{orcid.png}},
        Igor~Boiko\href{https://orcid.org/0000-0003-4978-614X}{\includegraphics[scale=0.75]{orcid.png}},~\IEEEmembership{Senior~Member,~IEEE,}
        and~Yahya~Zweiri\href{https://orcid.org/0000-0003-4331-7254}{\includegraphics[scale=0.75]{orcid.png}},~\IEEEmembership{Member,~IEEE}% <-this % stops a space

\thanks{

Manuscript received Month xx, 2xxx; revised Month xx, xxxx; accepted Month x, xxxx.
This work was supported by Khalifa University Grants CIRA-2020-082 and RC1-2018-KUCARS. (\textit{Corresponding author: Abdulla Ayyad})

A. Ayyad, P. Silva, M. Chehadeh, M.Wahbah, O. Hay, I. Boiko are with the Center for Autonomous Robotic Systems, Khalifa University, Abu Dhabi, United Arab Emirates (email: abdulla.ayyad@ku.ac.ae; pedro.silva@ku.ac.ae; mohamad.chehadeh@ku.ac.ae; mohamad.wahbah@ku.ac.ae; oussama.hay@ku.ac.ae; igor.boiko@ku.ac.ae).

Yahya Zweiri is with the Faculty of Science, Engineering and Computing, Kingston University London, London, SW15 3DW, UK. He is also with the Center for Autonomous Robotic Systems, Khalifa University, Abu Dhabi,
United Arab Emirates (e-mail: Y.Zweiri@kingston.ac.uk).

* Abdulla Ayyad, and Mohamad Chehadeh contributed equally to this work.
}}

% note the % following the last \IEEEmembership and also \thanks - 
% these prevent an unwanted space from occurring between the last author name
% and the end of the author line. i.e., if you had this:
% 
% \author{....lastname \thanks{...} \thanks{...} }
%                     ^------------^------------^----Do not want these spaces!
%
% a space would be appended to the last name and could cause every name on that
% line to be shifted left slightly. This is one of those "LaTeX things". For
% instance, "\textbf{A} \textbf{B}" will typeset as "A B" not "AB". To get
% "AB" then you have to do: "\textbf{A}\textbf{B}"
% \thanks is no different in this regard, so shield the last } of each \thanks
% that ends a line with a % and do not let a space in before the next \thanks.
% Spaces after \IEEEmembership other than the last one are OK (and needed) as
% you are supposed to have spaces between the names. For what it is worth,
% this is a minor point as most people would not even notice if the said evil
% space somehow managed to creep in.

% The paper headers
\markboth{IEEE Transactions on Control Systems Technology,~Vol.~14, No.~8, August~2015}%
{Shell \MakeLowercase{\textit{et al.}}: Bare Demo of IEEEtran.cls for IEEE Journals}
% The only time the second header will appear is for the odd numbered pages
% after the title page when using the twoside option.
% 
% *** Note that you probably will NOT want to include the author's ***
% *** name in the headers of peer review papers.                   ***
% You can use \ifCLASSOPTIONpeerreview for conditional compilation here if
% you desire.

% If you want to put a publisher's ID mark on the page you can do it like
% this:
%\IEEEpubid{0000--0000/00\$00.00~\copyright~2015 IEEE}
% Remember, if you use this you must call \IEEEpubidadjcol in the second
% column for its text to clear the IEEEpubid mark.

% use for special paper notices
%\IEEEspecialpapernotice{(Invited Paper)}

% make the title area
\maketitle

% As a general rule, do not put math, special symbols or citations
% in the abstract or keywords.
\begin{abstract}
Low cost real-time identification of multirotor unmanned aerial vehicle (UAV) dynamics is an active area of research supported by the surge in demand and emerging application domains. Such real-time identification capabilities shorten development time and cost, making UAVs' technology more accessible, and enable a wide variety of advanced applications. In this paper, we present a novel comprehensive approach, called DNN-MRFT, for real-time identification and tuning of multirotor UAVs using the Modified Relay Feedback Test (MRFT) and Deep Neural Networks (DNN). The main contribution is the development of a generalized framework for the application of DNN-MRFT to higher-order systems. One of the notable advantages of DNN-MRFT is the exact estimation of identified process gain, which mitigates the inaccuracies introduced due to the use of the describing function method in approximating the response of Lure's systems. A secondary contribution is a generalized controller based on DNN-MRFT that takes-off a UAV with unknown dynamics and identifies the inner loops dynamics in-flight.
Using the developed framework, DNN-MRFT is sequentially applied to the outer translational loops of the UAV utilizing in-flight results obtained for the inner attitude loops. DNN-MRFT takes on average 15 seconds to get the full knowledge of multirotor UAV dynamics and without any further tuning or calibration the UAV would be able to pass through a vertical window, and accurately follow trajectories achieving state-of-the-art performance. Such demonstrated accuracy, speed, and robustness of identification pushes the limits of state-of-the-art in real-time identification of UAVs. %250/250 words
\end{abstract}

% DNN-MRFT takes on average 15 seconds to get the full knowledge of multirotor UAV dynamics and was tested on multiple designs and sizes. The identification accuracy of DNN-MRFT is demonstrated by the ability of a UAV to pass through a vertical window without any further tuning or calibration.

% Note that keywords are not normally used for peerreview papers.

% For peer review papers, you can put extra information on the cover
% page as needed:
% \ifCLASSOPTIONpeerreview
% \begin{center} \bfseries EDICS Category: 3-BBND \end{center}
% \fi
%
% For peerreview papers, this IEEEtran command inserts a page break and
% creates the second title. It will be ignored for other modes.
\IEEEpeerreviewmaketitle
\nomenclature{$\aleph$}{Measurement noise.}

\nomenclature{$\beta$}{Modified relay feedback test phase parameter.}
\nomenclature{$\varphi_{d}$}{Distinguishing phase.}
\nomenclature{$PM$}{Phase margin.}
\nomenclature{$\Phi$}{Set of distinguishing phases arranged as a tree data structure.}
\nomenclature{$\Gamma$}{A mapping provided by a deep neural network.}
\nomenclature{$\tau$}{Generic time delay.}
\nomenclature{$\theta$}{Vehicle roll angle.}
\nomenclature{$\phi$}{Vehicle pitch angle.}
\nomenclature{$\psi$}{Vehicle yaw angle.}
\nomenclature{$C(s)$}{Generic controller in Laplace domain.}
\nomenclature{$C^*(s)$}{Optimal controller of a process.}
\nomenclature{$D$}{Generic domain of process parameters.}
\nomenclature{$\bar{D}$}{Generic discretized domain of process parameters.}
\nomenclature{$\mathcal{F}_B$}{Body attached reference frame.}
\nomenclature{$\mathcal{F}_I$}{Inertial reference frame.}
\nomenclature{$G(s)$}{Generic process in Laplace domain.}
\nomenclature{$h$}{Modified relay feedback test amplitude.}
\nomenclature{$Q$}{Integral squared error cost function.}
\nomenclature{$J_{ij}$}{Non-commutative joint cost between process \(G_i(s)\) and \(G_j(s)\).}
\nomenclature{$K$}{Generic static gain.}
\nomenclature{$N$}{Number of processes in certain discretized domain.}
\nomenclature{$N_d$}{Describing function of a non-linearity.}
\nomenclature{$u_0$}{Bias in a dynamical system.}
\nomenclature{$U(s)$}{Generic controller output in Laplace domain.}
\nomenclature{$R$}{Rotation matrix in the special orthogonal group \(SO(3)\).}
\nomenclature{$T$}{Generic time constant.}
\nomenclature{$X$}{Input matrix for deep neural network.}
\nomenclature{$C_{TW}$}{Peak-thrust to weight ratio.}
\nomenclature{$l_{mm}$}{Motor to motor distance, \(m\).}
\nomenclature{$I_{x}, I_y, I_z$}{Rotational inertial around \(x_B(+)\), \(y_B(+)\), \(z_B(+)\) axes, \(kg \cdot m^2\).}
\printnomenclature

\section{Introduction}
% The very first letter is a 2 line initial drop letter followed
% by the rest of the first word in caps.
% 
% form to use if the first word consists of a single letter:
% \IEEEPARstart{A}{demo} file is ....
% 
% form to use if you need the single drop letter followed by
% normal text (unknown if ever used by the IEEE):
% \IEEEPARstart{A}{}demo file is ....
% 
% Some journals put the first two words in caps:
% \IEEEPARstart{T}{his demo} file is ....
% 
% Here we have the typical use of a "T" for an initial drop letter
% and "HIS" in caps to complete the first word.
\IEEEPARstart{O}{line} controller parameters' estimation which accounts for unknown or changing process parameters has always been of interest in the controls community. Maintaining performance and safety are the main challenges that are often tackled in adaptive control research. A unified definition for adaptive control has always been a topic of discussion in the controls community, but we found the one from \cite{astrom1995adaptive} suitable and covers most relevant research. In \cite{astrom1995adaptive}, an adaptive controller is "a controller with adjustable parameters and a mechanism for adjusting the parameters". In this sense, adaptive controllers are of many different types and can extend to very complex formulations \cite{Kai2019}. In this paper we build on a novel technique \cite{ayyad2020real} that uses deep neural network (DNN) and the Modified Relay Feedback Test (MRFT) \cite{Boiko2013} to identify unknown process parameters. Specifically, in this work we investigate extending the approach suggested in \cite{ayyad2020real} to identify side motion dynamics of a symmetric multirotor vertical take-off and landing (VTOL) unmanned aerial vehicle (UAV) (in this document referred to simply as multirotor UAVs) which is under-actuated and has modeled process dynamics of relative degree five in addition to time delay. We show that our two-stage adaptive scheme can identify process parameters in real-time with high accuracy (first stage) and then suggest optimal controller gains based on the identified system parameters (second stage). We demonstrate that using our approach, a multirotor UAV can take-off without any pre-tuned controller gains and find the optimal controller parameters in-flight. To the best of our knowledge, this is the first adaptive controller that is capable of performing a takeoff and reach optimal controllers without initial stabilizing controller gains. 

Such demonstrated capability can be a game changer in the UAV industry as it shortens development time and cost, and expands the accessibility of UAV technology. For example, it benefits both the hobbyists community and enterprises that require custom UAV solutions by enabling safe and high-performance operation of custom built models in the shortest possible time. Additionally, the presented take-off and self-tuning approaches can be used in more advanced applications that requires real-time control gains adaptation while guaranteeing stability limits. A demonstration video of the presented approach applied to multiple multirotor UAVs that shows robustness in identification phase and high performance in the control phase can be found in \cite{paper_video}. 

\subsection{Relation to Existing Adaptive Control Approaches}
Adaptive control approaches are broad in nature and studying the relation of this work to all of them is not feasible. Many adaptive methods were applied to UAVs in simulation but they might suffer when applied experimentally due to over-simplified models or unrealistic adaptation gains \cite{Xu2018CLRC,MOFID20181,He2018adaptiveuav,Nadda2018adaptivesliding}. Thus we focus our study on adaptive approaches applied to UAVs experimentally. Also, we chose to extend our literature review to system identification methods that outputs an identified model in a form suitable for controller design as long as this identification was demonstrated experimentally. We chose to limit the scope of the literature review by excluding adaptive methods that deal with very specific cases; e.g. adapting to thrust coefficient change due to ground effect, change of lift force due to a propulsion fault, weight imbalance across a single axis, etc. 

One of the earliest approaches which demonstrated great success in UAV tuning is iterative learning control (ILC) \cite{Schoellig2012}. ILC tunes a feedforward law that compensates for repeatable model uncertainties. ILC requires a high number of experimental iterations and hence cannot adapt in real-time. Additionally, the feedforward compensation technique might suffer from severely degraded performance under unseen external disturbances or changes of model parameters. The approach presented in \cite{Berkenkamp2016} utilizes Gaussian processes with Bayesian optimization to learn feedback control parameters for the translational control loops. This approach requires a lot of iterations to converge and hence cannot be applied to real-time applications. Deep model-based reinforcement learning (RL) was used to adapt a RL based control policy from experimentation \cite{lambert2019low}. This method requires excessive experimental data, is computationally expensive, and does not provide stability guarantees. Another RL method that relies on training an RL agent by closing the simulation to reality (S2R) gap is presented in \cite{molchanov2019sim}. In the work of \cite{molchanov2019sim}, the use of domain randomization technique and detailed models helped reducing the S2R gap such that generalized policies that are directly transferable to platforms of different sizes were successfully generated. State space sampling exploration techniques through deep learning \cite{li2017deep}, and apprenticeship learning \cite{abbeel2010autonomous} were used to fine-tune and improve the performance of outer loop controllers. These techniques require an abundance of experimental data and offline computation.

Other tuning approaches based on relay methods have been applied in practice. Recent work in \cite{Chehadeh2019} shows near-optimal attitude loops controller tuning based on MRFT. Though this tuning method can run in real-time, its tuning performance degrades in the presence of biases in the system \cite{ayyad2020real}. Another relay based tuning method that uses relay feedback test (RFT) in a cascaded arrangement is presented in \cite{poksawat2016automatic}. The tuning assumes first order plus time delay (FOPTD) model and was only performed on a testbed. Heuristic and model-free approaches were widely investigated in literature  \cite{giernacki2019iterative,giernacki2019real,tesch2016pitch,howard2017platform} but their tuning time is generally large (a few minutes at least) and there is no guarantee of optimality of the achieved controller gains. Few other methods of UAV tuning are based on experimental system identification. In \cite{Du2014ga_identification}, frequency-domain identification using an adaptive genetic algorithm was performed on an unmanned helicopter. The identification method requires a fair amount of flight data, which in turn requires a pre-processing stage that includes human expertise. Similar drawbacks are present in the approach used by \cite{Lin2011ident,ALSHARMAN2018} where UAV models based on fuzzified eigensystem realization algorithm were identified. 

A common limitation of all reviewed approaches is that they require a stabilized system to begin with. This is usually done through an extensive trial and error process or initial rough tuning based on pre-measured physical parameters. This leads to a prolonged development time and increased cost especially for larger UAVs. Also, most of the presented approaches can be exclusively used either to adapt attitude and attitude rate loops gains (inner loops), or outer loops gains. The literature lacks a unified robust approach for tuning of the inner and outer loops of multirotor UAVs. Another limitation specific to data-driven approaches like ILC, state-space sampling approaches, and other identification methods widely adopted in literature \cite{McElveen1992ident,Nevaranta2016dft,Erdogan2020ident} is that tuning performance is dependent on how data is generated. Data generation for these adaptation techniques has its own complexities and requires an expert human to perform.

Our proposed approach uses MRFT, which can be considered as an extension of the widely used RFT, to excite a certain system response. This system response is fed to a DNN that is able to infer system parameters. Therefore, we refer to the approach presented in this paper by DNN-MRFT. DNN-MRFT is the most appealing approach compared to the other relevant adaptive approaches described in literature due to its stability guarantees, its minimal data requirements, and its computational efficiency which enables it application in real-time. DNN-MRFT provides additional benefit that results in accurate identification of model parameters, permitting the design of controllers other than PID. Thus DNN-MRFT can be also considered as a system identification method. The DNN is only trained on simulation data which greatly simplifies the identification algorithm design process.

\subsection{Contributions}

DNN-MRFT provides a unified approach for the identification of a linear system's parameters. It was first introduced and applied to a second order plus integrator plus time delay (SOIPTD) system depicting multirotor UAV attitude dynamics \cite{ayyad2020real}. The main contribution of the present paper is to investigate  the viability of using the DNN-MRFT approach to multirotor UAV side motion dynamics. These dynamics can be formulated as a fifth order system with time delay, which requires a different treatment than SOIPTD model presented in our previous work \cite{ayyad2020real}. Existing DNN-MRFT formulation is not suited for the identification of such higher order systems as MRFT can only produce periodic oscillations corresponding to the second and third quadrants of the complex plane. Harmonic balance (HB) anticipates oscillations with amplitudes that can reach up to tens of meters when running conventional DNN-MRFT on processes considered in Table \ref{tab:parameters_range}, which clearly indicates impracticality of the conventional formulation. The solution to this challenge is in the extension of DNN-MRFT to higher order dynamics in a hierarchical fashion where identified inner loop dynamics, including inner loop controller, are considered in the identification of outer loop dynamics. This hierarchical identification approach through DNN-MRFT can be repeated to generalize for any higher order systems. As a result of the DNN-MRFT approach, it is possible to exactly identify open-loop system gain utilizing simulated knowledge of the considered systems. This mitigates amplitude determination inaccuracy due to the Describing Function (DF) method's low pass filtering assumption. In literature, the Locus of the Perturbed Relay System (LPRS) \cite{boiko1999lprs} was suggested as an exact description of discontinuous systems. The proposed amplitude scaling technique in this paper is simpler, and can be directly used in controller tuning. A secondary contribution of this paper is the development of an algorithm based on the DNN-MRFT hierarchical identification approach that allows a multirotor UAV to takeoff without any initial controller parameters and perform identification and tuning safely for all control loops. This takeoff algorithm is specific to multirotor UAVs and was tested experimentally on multiple sizes and designs of multirotor UAVs. The overall performance of UAV identification is demonstrated by two experiments. First we achieve trajectory tracking performance on par with state-of-the-art. Second, using the DNN-MRFT identification results, the multirotor UAV can pass through a vertical narrow window without the need of fine-tuning or any other sort of human expert input to the controller structure or gains. To the best of our knowledge, this is the only adaptive scheme that can take-off a multirotor UAV with zero initial gains and achieve a feedback controller that can perform such aggressive maneuvers in a completely autonomous manner.

\subsection{Paper Outline}
This paper is organized as follows: aspects related to dynamics modeling and relevant assumptions are discussed in Section \ref{sec:mdl_dyn}. Overall theoretical framework of DNN-MRFT identification including the design of MRFT parameters through finding the \textit{distinguishing phase} and relevant stability considerations are discussed in Section \ref{sec:mrft_ident}. The process of discretizing the model parameter space into a finite set of representative processes is described in  Section \ref{sec:class_outputs}. The DNN model development and the generation of training data through simulation is discussed in Section \ref{sec:data_gen}. A modified method for finding exact system gain that mitigates the DF approximation is presented in Section \ref{sec:gain_correction}. The design of an algorithm that can perform safe take-off, identification, and tuning of optimal controllers for UAVs is shown in Section \ref{sec:take_off_cont}. Finally, extensive simulation and experimental results which demonstrates state-of-the-art performance and adaptation robustness are presented in Section \ref{sec:exp_results}.

% needed in second column of first page if using \IEEEpubid
%\IEEEpubidadjcol

\section{Modelling of Dynamics}
\label{sec:mdl_dyn}
In this work, we define the inertial frame \(\mathcal{F}_I\) to be earth-fixed right-handed reference frame with \(z_I(+)\) pointing upwards. The right-handed body reference frame \(\mathcal{F}_B\) is attached to the multirotor UAV center of mass, with \(z_B(+)\) perpendicular to the body upper surface, and is always aligned with its attitude and heading angles. A rotation matrix used to transform between reference frames is denoted by \({}^T_SR\), where \(T\) is the target reference frame and \(S\) is the source one.

\begin{figure}[t]
\includegraphics[width=0.49\textwidth]{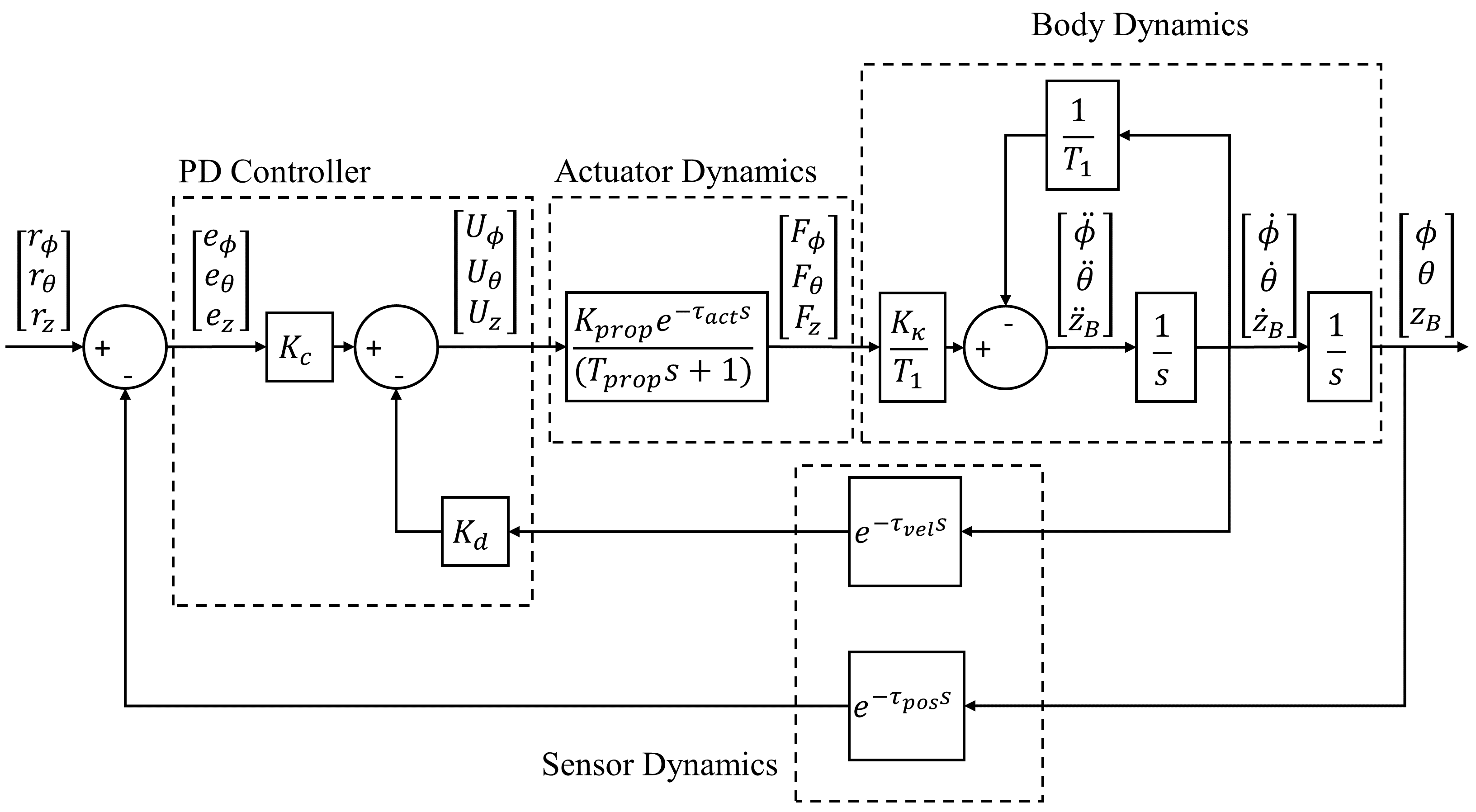} 
\caption{The generic model used for attitude and altitude dynamics under PD feedback control.}
\label{fig:inner_loop_model}
\end{figure}

\subsection{Modeling of Attitude and Altitude Dynamics}
The approach of DNN-MRFT was previously applied to a SOIPTD system and demonstrated in accurate identification results \cite{ayyad2020real}. Attitude and altitude share the same model structure but each have different model parameters. Attitude and altitude loops are modeled as \cite{Chehadeh2019}:
\begin{equation}\label{eq_attitude_model_topdt}
G_{inner}(s)=\frac{K_{eq}e^{-\tau s}}{s(T_{prop}s+1)(T_{1}s+1)}
\end{equation}
A more detailed representation of these dynamics with PD feedback control can be seen in Fig. \ref{fig:inner_loop_model}. The linear dynamics in Eq. \eqref{eq_attitude_model_topdt} relate motor commands sent by the flight controller to the observed roll, pitch, or altitude. Note that the time delay in the numerator represents the overall time delay in the system which consists of electronic speed controller (ESC), processor, communication and sensor delays. The nonlinearity of the system is mainly exhibited by the change in the value of the parameter \(T_{1}\) as a function of rotational velocity (for attitude dynamics), or translational velocity (for altitude dynamics) representing nonlinear drag dynamics. The assumption that such a drag effect, caused by air inflow, blade flapping, and body drag, can be considered constant works well in practice and was analyzed in detail by \cite{hoffman2007,pounds2010}. \(K_{eq}\) represents the overall open-loop gain of the system and is composed of two gains; \(K_{prop}\) which represents propulsion system gain, and \(K_{\kappa}\) which represents a static mapping that takes into account motors' configuration and geometric properties to produce specific thrust and torque quantities. Propulsion systems, consisting of electronic speed controllers ESCs and motors, are assumed to provide linear response of thrust function of ESC command; and hence, \(K_{eq}\) can be considered constant. From bench propulsion system tests similar to the ones performed in \cite{Cheron2010}, it can be concluded that \(T_{prop}\) is constant across the whole operating range except when the rotational speed of the motor is very low. In practice, we avoid operating in this non-linear range by enforcing appropriate minimum motor command. Additionally, network communication and processing delays are almost constant (i.e. have small variance in delay value), permitting us to consider the time delay \(\tau\) as a constant. The considered attitude and altitude dynamics are subject to measurement noise \(\aleph\) and forced bias \(u_0\) due to external disturbances such as gravity, sensor bias, unmatched propulsion thrust, or model asymmetry. 

The coupled dynamics of rotational motion due to the gyroscopic effect is assumed to be negligible. This is because in the operational limits we are interested in, the torques generated due to gyroscopic effects are considerably smaller than the torques contributed by other dynamics of the system \cite{Chehadeh2019}. Thus the assumption of single input single output (SISO) system dynamics for every rotational control loop is valid. Coupling of rotational dynamics can also occur due to other reasons like sensor misplacement, asymmetric center of mass, etc. Care was taken to minimize such effects when preparing the experimental setup.

\begin{figure}[t]
\includegraphics[width=0.49\textwidth]{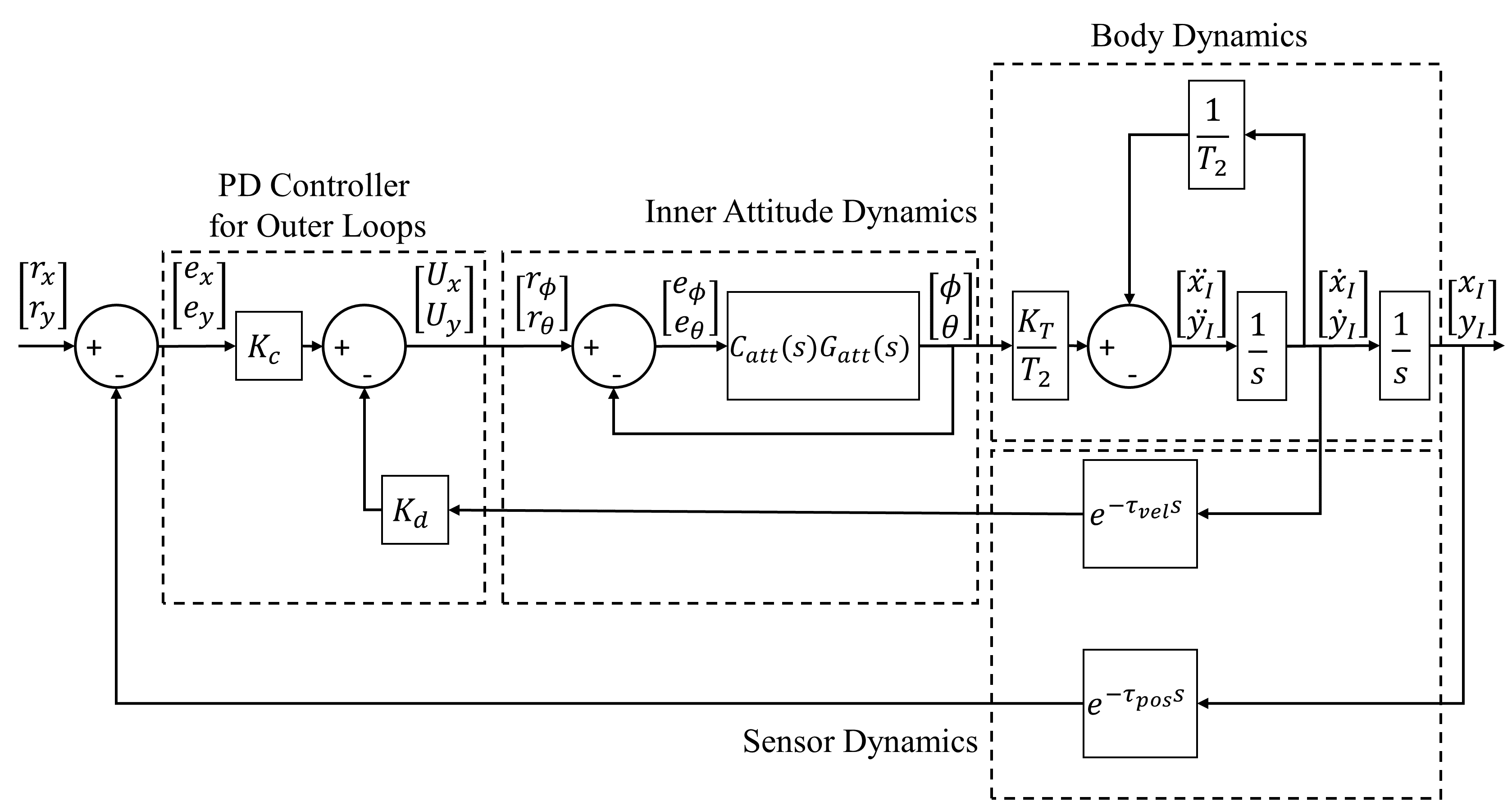} 
\caption{The generic model used for side motion dynamics (\(x_I\) and \(y_I\)) under PD feedback control. \(T_2\) represents drag term due to translational velocity component, and \(K_T\) represents the approximated small angle conversion coefficient to specific thrust.}
\label{fig:outer_loop_model}
\end{figure}

\subsection{Modeling of Side Motion Dynamics}
Multirotor UAVs are underactuated due to the fact that movements in the \(x_B(+)\) and \(y_B(+)\) due to actuator action are not possible. But side movements in the inertial frame \(\mathcal{F}_I\) are possible and can be approximated to be linear for small attitude angles. The linearized side motion dynamics are given by:
\begin{equation}\label{eq_side_motion_model_foptd}
G_{outer}(s)=\frac{K_{eq,att}K_{eq,side}e^{-(\tau_{att}+\tau_{side}) s}}{s^2(T_{prop}s+1)(T_{1}s+1)(T_{2}s+1)}
\end{equation}
Here we assume a linear drag term \(T_{2}\) that describes air resistance acting on the body of the multirotor UAV frame due to translational motion. The assumption of linearity works well in practice for small angles due to the fact that translational drag terms remain similar, and altitude loss due to thrust vector change is negligible (i.e. assume a nominal value of thrust provided by the motors). Overall system dynamics with feedback control design for small angles are shown in Fig. \ref{fig:outer_loop_model}. Note that the attitude (inner loop) model parameters directly affect the performance of the outer loop dynamics.

In cases where aggressive maneuvers require large rotation angles, we present a different treatment for the generation of reference attitude and thrust. This is achieved by utilizing the controller structure presented in \cite{Mellinger2011}, where the differential flatness properties of multirotor UAVs are exploited such that the outer loop controller outputs a desired inertial force vector \(\vec{F}_{des}\) instead of directly producing reference rotation angles and thrust. The feedback control structure of \cite{Mellinger2011} neglects actuator and inner-loop dynamics, nonetheless, the identification and tuning approach presented in this paper takes these dynamics into account. The inertial force vector is related to the UAV's attitude and thrust by the following equation:
\begin{equation}
    \vec{F} = \begin{pmatrix}
    F_x \\
    F_y \\
    F_z
    \end{pmatrix} = {}^I_BR
    \begin{pmatrix}
    0 \\
    0 \\
    U_{z}
    \end{pmatrix}
    -
    m
    \begin{pmatrix}
    0 \\
    0 \\
    g
    \end{pmatrix}
    \label{eq_force_model}
\end{equation}

%we consider \(\vec{F}_{des}\) to include a bias term that cancels the gravitational effect, and use it to
Following \eqref{eq_force_model}, we use \(\vec{F}_{des}\) to solve for the desired UAV attitude and thrust. We define a reference frame \(\mathcal{F}_{B^*}\) that represents the desired UAV attitude as follows:
\begin{equation}
    z_{B^{*}}(+) = \frac{\vec{F}_{des}}{\norm{\vec{F}_{des}}}
    \label{eq_target_attitude_z}
\end{equation}
\begin{equation}
    y_{B^{*}}(+) = \frac{z_{B^{*}} \times \begin{bmatrix} 
    \cos{r_{\psi}} & \sin{r_{\psi}} & 0
    \end{bmatrix}^T}{\norm{z_{B^{*}} \times \begin{bmatrix} 
    \cos{r_{\psi}} & \sin{r_{\psi}} & 0
    \end{bmatrix}^T}}
    \label{eq_target_attitude_y}
\end{equation}
\begin{equation}
    x_{B^{*}}(+) = y_{B^{*}}(+) \times z_{B^{*}}(+)
    \label{eq_target_attitude_x}
\end{equation}
where \(r_{\psi}\) is the reference yaw angle.

The roll, pitch, and yaw components of the orientation error successively denoted by \(e_{\phi}\), \(e_{\theta}\), and \(e_{\psi}\) can then be computed as:
\begin{equation}
    \begin{pmatrix}
    e_{\phi} \\
    e_{\theta} \\
    e_{\psi} \\
    \end{pmatrix}
    =  \mathcal{E}({}^{B^{*}}_IR \; {}^I_BR)
    \label{eq_angles_large}
\end{equation}
where \(\mathcal{E}\) is a function that converts a rotation matrix in \(SO(3)\) to the corresponding rotation vector in \(\mathrm{R}^{3}\). 

Finally, we solve for \(U_{z}\) as follows:
\begin{equation}
    U_{z} =  \vec{F}_{des} \cdot	z_B(+)
    \label{eq_thrust_projection}
\end{equation}

\subsection{Yaw Dynamics}
Rotation around \(z_B\) axis results in change of the yaw angle. Yaw has second order dynamics and is given by:
\begin{equation}\label{eq_yaw_soptd}
G_{yaw}(s)=\frac{K_{eq}e^{-\tau_{att} s}}{s(T_{prop}s+1)}
\end{equation}
Because yaw controller is easy to tune due to the small delay value and the presence of full state measurements, we assume that a controller with satisfactory performance exists prior to the flight.

\subsection{Bounds of Considered Model Parameters}
The identification method presented in this paper requires the considered model parameters to be bounded. This would limit the amount of data and labels to be handled by the DNN classifier presented later. In this work, we consider commonly used multirotor UAV designs ranging from small racing quadrotors to larger multirotors with take-off weight of up to approximately 50Kgs. The selection of the parameters domain was based both on experimental findings of previous work in the literature \cite{Bouabdallah2004,Chehadeh2019,pounds2010,Cheron2010}, in addition to modeling equations like those discussed in \cite{Chehadeh2019,pounds2010,hoffman2007,prouty2002}. It is worth noting that the identification performance is not sensitive to the selection of the parameters' range, rather, the selection of parameter bounds can be safely expanded to include UAV designs beyond the specified ranges. Such expansion would be at the cost of increased simulation and DNN training times. The selected bounds of model parameters for the considered control loops can be found in Table \ref{tab:parameters_range}.

\renewcommand{\arraystretch}{2}
\begin{table}[h]
\caption{The model parameters' ranges for all the feedback loops considered in this paper}
    \centering
 \begin{tabular}{|p{0.13\textwidth}|p{0.315\textwidth}|} 
 \hline
 Parameters & Domain \\ [0.5ex] 
 \hline
 Attitude & 
 \(D_{att}:=\{(T_{prop},T_{1},\tau_{att}):0.015\leq T_{prop}\leq 0.3,\,0.2\leq T_{1}\leq 2,\,0.0005\leq \tau_{att}\leq 0.03\}\) \\ 
\hline
 Altitude & 
 \(D_{alt}:=\{(T_{prop},T_{1},\tau_{alt}):0.015\leq T_{prop}\leq 0.3,\,0.2\leq T_{1}\leq 2,\,0.0005\leq \tau_{alt}\leq 0.15\}\) \\ 
\hline
 Side & 
 \(D_{side}:=\{(T_{2},\tau_{side}):0.2\leq T_{2}\leq 6,\,0.0005\leq \tau_{side}\leq 0.15\}\) \\
 \hline
 
 Peak Thrust-to-Weight Ratio & 
 \(C_{TW} \in [1.5,5]\) \\
 \hline
 Motor-to-motor distance (\(m\))  & 
 \(l_{mm} \in [0.15,1.5]\) \\
 \hline
 Moment of Inertia (\(kg\cdot m^2\)) & 
 \(\{I_x, I_y\} \in [1.6\times 10^{-3},2.25]\) \\
 \hline
\end{tabular}
\label{tab:parameters_range}
\end{table}

\section{MRFT and Identification Approach}
\label{sec:mrft_ident}

\subsection{The Modified Relay Feedback Test}
DNN-MRFT relies on exciting certain system response using MRFT as a controller. MRFT is an algorithm that can excite self-sustained oscillations at a specific phase \(\varphi\), and is realized by the following equation \cite{Boiko2013}:
\begin{multline}\label{eq_mrft_algorithm}
u_M(t)=\\
\left\{
\begin{array}[r]{l l}
h\; &:\; e(t) \geq b_1\; \lor\; (e(t) > -b_2 \;\land\; u_M(t-) = \;\;\, h)\\
-h\; &:\; e(t) \leq -b_2 \;\lor\; (e(t) < b_1 \;\land\; u_M(t-) = -h)
\end{array}
\right.
\end{multline}
where \(b_1=-\beta e_{min}\) and \(b_2=\beta e_{max}\). \(e_{max}>0\) and \(e_{min}<0\) are respectively the last maximum and minimum values of the error signal after crossing the zero level; and \(u_M(t-)=lim_{\epsilon\rightarrow0^+ }u_M(t-\epsilon)\) is the previous control signal. Prior to the start of MRFT, the maximum and minimum error values are set as: \(e_{max}=e_{min}=0\). \(\beta\) is a constant parameter that dictates the phase of the excited oscillations as:
\begin{equation}
\label{eq:phase_crossover_beta}
    \varphi = \arcsin{(\beta)}
\end{equation}

Using the DF method, it could be shown that the MRFT achieves oscillations at a specified phase angle by satisfying the HB equation \cite{atherton1975}:
\begin{equation}\label{eq_hb}
N_d(a_0)G(j\Omega_0)=-1
\end{equation}
The DF of MRFT is presented in \cite{Boiko2013} as:

\begin{equation}\label{eq_mrft_df}
N_d(a_0)=\frac{4h}{\pi a_0}(\sqrt{1-\beta^{2}}-j\beta)
\end{equation}

The DF method provides an approximate solution that is valid only if \(G(s)\) has sufficient low pass filtering properties. It is worth mentioning that the MRFT control signal \(u_M(t)\) has a phase lead relative to the error signal \(e(t)\) in the case of \(\beta<0\), and lags in the case of \(\beta>0\). The MRFT DF intersects the Nyquist plot in the second quadrant for \(\beta<0\); while this intersection occurs in the third quadrant when \(\beta>0\). The Relay Feedback Test (RFT) \cite{Astrom1984} could be thought of as a special case of the MRFT algorithm where \(\beta=0\).

\subsection{The Distinguishing Phase}
\label{sec:dist_phase}
The idea of distinguishing phase is based on the supposition that the optimal phase angle at which the test oscillations are generated and which is obtained through the design of optimal tuning rules \cite{Boiko2014,Boiko2013,Chehadeh2019} would reveal the most distinguishing characteristics of the considered processes domain. In a previous work \cite{ayyad2020real}, we showed that for an LTI system \(G(s)\) with known model structure and unknown set of bounded model parameters \(D\), there exists a distinguishing phase \(\varphi_{d}\) at which the characteristics of the self-excited oscillations induced by the MRFT can be used to identify the corresponding processes in \(D\). The distinguishing phase \(\varphi_{d}\) can be determined by the process of designing optimal non-parametric tuning rules as outlined in \cite{Boiko2013,Boiko2014}. Note that MRFT parameter \(\beta\) is related to the distinguishing phase by Eq. \eqref{eq:phase_crossover_beta}. Algorithm \ref{alg:dist_phase} summarizes the steps taken to find the value of \(\varphi_{d}\).

\begin{algorithm}[h]
\SetAlgoLined
\begin{flushleft}
        \textbf{INPUT:} (G(s),\(D\)) - Model Structure, Parameters Domain \\
        \textbf{OUTPUT:} \(\varphi_{d}\) - Distinguishing Phase
\end{flushleft}
%\KwResult{ }
\begin{algorithmic}[1]
\STATE Discretize the desired parameters subspace \(D\) to obtain \(\bar{D}\)\;
\STATE Select phase margin or gain margin tuning specifications\;
\STATE Find the set of locally optimal tuning rules \(\Delta\) for every process in \(\bar{D}\)\;
\STATE Apply every optimal tuning rule in \(\Delta\) to all other processes in \(\bar{D}\) and get the set \(\Sigma\) corresponding to the value of the worst performance deterioration of every process in \(\bar{D}\) due to the application of the non-optimal tuning rule \;
\STATE Select the tuning rule from \(\Delta\) that corresponds to the least worst deterioration value from \(\Sigma\) as the globally optimum tuning rule \(\Delta^*\)\;
\STATE Compute \(\varphi_d\) from \(\beta \in \Delta^*\)\; 
\end{algorithmic}

 \caption{Finding distinguishing phase through optimal non-parametric tuning rules design}
 \label{alg:dist_phase}
\end{algorithm}
\begin{comment}
It was shown in \cite{ayyad2020real} that a single steady-state oscillation excited by MRFT at \(\varphi_d\) is sufficient to identify unknown model parameters.
\end{comment}

%Ayyad notes:
\begin{comment}
We need to mention the below main ideas:
1- The outer loop identification utilizes information from inner loop.
2- THe outer loop is identified assuming an optimal controller is acting on the inner loop.
\end{comment}

\subsection{Identification Approach}
The MRFT can excite stable periodic oscillations only in the second and third quadrants of the complex plane. As a result, for high relative degree systems, the generated test oscillations have values of \(a_0\) and \(\Omega_0\) from Eq. \eqref{eq_hb} that may not be practically useful. This is a reflection of the importance of using a cascade controller arrangement, which would in turn require to organize the MRFT tests in each loop separately to tune each controller. It would also eliminate the indicated problem of the test oscillations possibly being of low frequency and high amplitude. For this reason, the considered high order LTI system must be split into a composition set \(G_{HO}:=\{G_1,G_2,...,G_M\}\) where \(G_1\) represents the sub-system with smallest relative degree with respect to the control command. The iterative design method required to generate the feedback structure and hence required set of distinguishing phases for higher order LTI processes can be found in Algorithm \ref{alg:dist_phase_ho}. Note that Algorithm \ref{alg:dist_phase_ho} assumes that inner feedback loops are observable and controllable.
\begin{comment}
\begin{algorithm}[h]
\caption{Generating cascaded distinguishing phases}
\label{alg:dist_phase_ho}
\SetAlgoLined
\begin{flushleft}
        \textbf{INPUT:} \((G_{HO}(s),\bar{D})\) - Model Structure, Discretized Parameters Domain \\
        \textbf{OUTPUT:}  \(\Phi,G_{res}\) - Set of Distinguishing Phases, Resulting Feedback Structure
\end{flushleft}
\begin{algorithmic}[1]
\STATE Split \(G_{HO}\) to \(\{G_1,G_2,...,G_M\}\) based on M measurable or observable outputs with unity open loop gain \label{alg:dist_phase_ho:g_ho}
\STATE \(G_{res} \gets 1\)
\FOR{i=1,...,M}
\STATE \(G_{res} \gets G_{res}G_i\)
\IF {i=M}
\STATE \(\varphi_{d,i} \gets GetDistinguishingPhase(G_{res},D)\) from Algorithm \ref{alg:dist_phase}
\ELSE
\STATE \(\varphi_{d,i+1} \gets GetDistinguishingPhase(G_{i+1}G_{res},D)\) from Algorithm \ref{alg:dist_phase}
\ENDIF
\STATE \((a_0,\Omega_0) \gets MRFT(G_{i+1}G_{res},\varphi_d)\) from Eq. \eqref{eq_mrft_algorithm} 
\IF {\(Re\{N_d(a_0)G_{i+1}(j\Omega_0)\}\geq0\) \textbf{or} Impractical \(a_0\) or \(\Omega_0\) values \textbf{or} i=M} \label{alg:dist_phase_ho:cond}
\STATE \(C^* \gets\) Find optimal controller of \(G_{res}\)
\STATE \(G_{res} \gets\) Feedback(\(C^*\),\(G_{res}\))
\STATE \(\Phi \gets \Phi \cup \{\varphi_{d,i}\}\)
\ENDIF
\ENDFOR
\end{algorithmic}

\end{algorithm}
\end{comment}

\begin{algorithm}[h]
\caption{Generating cascaded feedback structure}
\label{alg:dist_phase_ho}
\SetAlgoLined
\begin{flushleft}
        \textbf{INPUT:} \((G_{HO}(s),D)\) - Model Structure, Parameters Domain \\
        \textbf{OUTPUT:}  \(G_{FB}\) - Resulting Feedback Structure
\end{flushleft}
\begin{algorithmic}[1]
\STATE Split \(G_{HO}\) to \(\{G_1,G_2,...,G_M\}\) based on M observable outputs with unity open loop gain \label{alg:dist_phase_ho:g_ho}
\STATE \(G_{res} \gets 1\)
\FOR{i=1,...,M-1}
\STATE \(G_{res} \gets G_{res}G_i\)
\STATE \(\bm{V_{i}} \gets\) All processes at the vertices of \(D\) parameters in \(G_{res}\)

\STATE \((a_0,\Omega_0) \gets MRFT(G_{i+1}\bm{V_{i}},\varphi_d)\) from Eq. \eqref{eq_mrft_algorithm}
\IF {\(Re\{N_d(a_0)G_{i+1}(j\Omega_0)\bm{V_{i}}(j\Omega_0)\}\geq0\) \textbf{or} Impractical \(a_0\) or \(\Omega_0\) values} \label{alg:dist_phase_ho:cond}

\STATE \(C^* \gets\) Find optimal controller of \(G_{res}\)
\STATE \(G_{res} \gets\) Feedback(\(C^*\),\(G_{res}\))
\ENDIF
\ENDFOR
\STATE \(G_{res} \gets G_{res}G_M\)
\STATE \(C^* \gets\) Find optimal controller of \(G_{res}\)
\STATE \(G_{FB} \gets\) Feedback(\(C^*\),\(G_{res}\))
\end{algorithmic}

\end{algorithm}

\begin{figure}[t]
\includegraphics[width=0.49\textwidth]{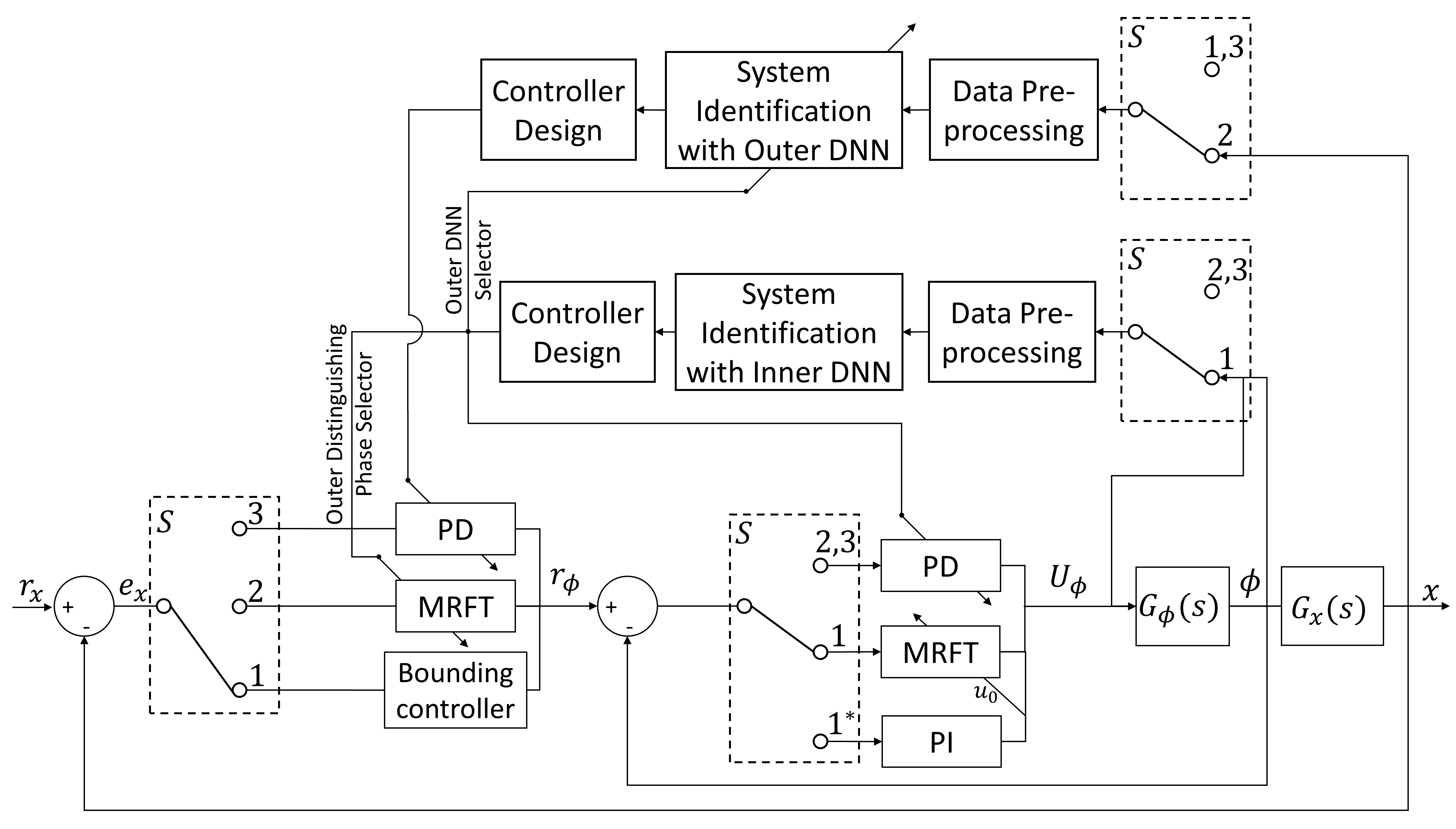} 
\caption{The identification scheme used for multirotor UAV side motion dynamics. Only inner part of the identification is applicable to altitude dynamics identification. Note that there is only one switch \(S\) in this identification scheme. The identification starts with \(S\) at \(1\) (and at \(1^*\) for altitude). Once enough data are pre-processed from steady-state MRFT response of inner loops, system identification is performed by the inner loops DNN. Once the inner loop systems are identified, an appropriate controller for each control loop is designed and \(S\) switches to \(2\). Note that the outer DNN structure and weights, and the MRFT \(\beta\) parameter are all selected based on the identified inner loop model parameters and designed controller. Once enough data are pre-processed from steady-state MRFT response of outer loops, system identification is performed by the outer loops DNN which is immediately followed by controller tuning. \(S\) switches to \(3\) and the system is controlled optimally.}
\label{fig:identification_scheme}
\end{figure}

For the particular case presented in this paper, we obtain two cascaded feedback loops by applying Algorithm \ref{alg:dist_phase_ho} to the side motion model in Eq. \eqref{eq_side_motion_model_foptd}. From Algorithm \ref{alg:dist_phase_ho} line \ref{alg:dist_phase_ho:g_ho} we get:
\begin{equation}\label{eq_side_motion_ho_alg}
\begin{split}
G_{HO}(s)\rightarrow \{G_1,G_2,G_3,G_4\}= \\
\{\frac{e^{-\tau_{att} s}}{(T_{body}s+1)(T_{prop}s+1)},\frac{1}{s},\frac{e^{-\tau_{side} s}}{(T_{side}s+1)},\frac{1}{s}\}
\end{split}
\end{equation}
Where \(G_1\) represents attitude rate dynamics, \(G_1G_2\) represents attitude dynamics given in Eq. \eqref{eq_attitude_model_topdt}, \(G_1G_2G_3\) represents side motion velocity dynamics, and \(G_{tot}=G_1G_2G_3G_4\) represents side motion dynamics given in Eq. \eqref{eq_side_motion_model_foptd}. The condition in Algorithm \ref{alg:dist_phase_ho} at line \ref{alg:dist_phase_ho:cond} is met only when \(i={2}\) for multirotor UAV side motion case which will result in two cascaded feedback loops shown in Fig. \ref{fig:outer_loop_model}. Also, this means that we will end up with a set of distinguishing phases; one distinguishing phase to reveal the inner loop attitude dynamics and a distinguishing phase for every process in \(\bar{D}_{att}\) to reveal outer loop position dynamics. Note that the distinguishing phase of the particular outer loop system depends on the inner closed-loop dynamics. Thus prior to outer loop identification, the parameters of the inner loop dynamics, and the optimal controller for the inner loop have to be identified first. The overall identification scheme used in this paper is shown in Fig. \ref{fig:identification_scheme}. It is important to note that this approach is generic and can be applied to higher order LTI systems as long as the distinguishing phase corresponds to second or third quadrants in the complex plane.

\subsection{Stability Considerations}
Stability aspects of the DNN-MRFT approach can be divided into two categories. First category includes stability of the periodic oscillations generated by MRFT when the system is in the identification phase. Existence of stable periodic solutions will be proved in this section using the DF method and Loeb\'s criterion \cite{loeb1956recent}. Other aspects related to stability in the sense specific to UAVs like boundedness of bias and oscillation's amplitude are discussed in Section \ref{sec:take_off_cont}. The second aspect of stability is the guarantee of system stability upon the controller parameters selection by the DNN classifier. Similar to the original formulation of the MRFT and coordinated test and tuning \cite{Boiko2013}, in which the stability is guaranteed with a specified gain or phase margins; the current approach guarantees stability for the class of dynamic systems being considered. Monte-Carlo sampling is used to demonstrate this aspect of stability as discussed in Section \ref{subsubsec:sim_results}.

Loeb\'s criterion provides necessary condition for asymptotic orbital stability \cite{loeb1956recent}. We follow Loeb criterion formulation similar to the one used in \cite{atherton1975,aguilar2015self}. Let a perturbation in amplitude denoted by \(\Delta a\) cause a change in the Laplace variable \(\Delta s=\Delta \sigma+j\Delta \Omega\) such that HB still holds:
\begin{equation}
    N(a_0+\Delta a)G(j\Omega_0 + \Delta \sigma+j\Delta \Omega)=-1
\end{equation}
Using fundamental theorem of calculus and differentiation chain rule we rearrange to obtain:
\begin{equation}
    (N(a_0)+\frac{\partial N}{\partial a}\Delta a)(G(j\Omega_0)+\frac{\partial G}{\partial \sigma}\frac{d\sigma}{da}\Delta a+\frac{\partial G}{\partial \Omega}\frac{d\Omega}{da}\Delta a)=-1
\end{equation}
We cancel resultant \(N(a_0)G(j\Omega_0)\) and \((\Delta a)^2\) terms and then divide all terms by \(N(a_0)G(j\Omega_0)\). Using the property \(\frac{\partial G}{\partial \sigma}=-j\frac{\partial G}{\partial \Omega}\) we obtain:
\begin{equation}
    \frac{\partial \ln{N}}{\partial a}-j\frac{\partial \ln{G}}{\partial w}\frac{d\sigma}{da}+\frac{\partial \ln{G}}{\partial w}\frac{dw}{da}=0
\end{equation}
By using the property \(\ln{z}=\ln{|z|}+j\Arg{z}\) we obtain:
\begin{equation}
    (\frac{\partial \ln{|G|}}{\partial \Omega}+j\frac{\partial \Arg{G}}{\partial \Omega})(\frac{d\Omega}{da}-j\frac{d\sigma}{da})=-\frac{d\ln{|N|}}{da}-j\frac{d \Arg{N}}{da}
\end{equation}
By multiplying the real part of the above equation with \(\frac{\partial \Arg{G}}{\partial \Omega}\), and the imaginary part with \(\frac{\partial \ln{|G|}}{\partial \Omega}\) and rearranging, we get the following equation:
\begin{equation}
    \frac{d\sigma}{da}=\frac{-\frac{d \ln{|N|}}{da}\frac{\partial \Arg{G}}{\partial \Omega}+\frac{d\Arg{N}}{da}\frac{\partial \ln{|G|}}{\partial \Omega}}{(\frac{\partial \ln{|G|}}{\partial \Omega})^2+(\frac{\partial \Arg{G}}{\partial \Omega})^2}
\end{equation}
For asymptotic orbital stability the condition \(\frac{d\sigma}{da}<0\) is necessary. Given the MRFT describing function presented in Eq. \eqref{eq_mrft_df}, the second term in the numerator of the above equation is zero. Also, considering the positive denominator in the above equation we can state the conditions of orbital stability of MRFT oscillations as:
\begin{equation}
\label{eq_loeb_criterion}
    -\frac{d \ln{|N|}}{da}\frac{\partial \Arg{G}}{\partial \Omega}<0
\end{equation}
Considering the DF of the MRFT we have:
\begin{equation}
    -\frac{d \ln{|N|}}{da}=\frac{-1}{|N|}\frac{d|N|}{da}=\frac{1}{a}
\end{equation}
which is always positive. Then the term \(\frac{\partial \Arg{G}}{\partial \Omega}\) needs to be strictly negative for at least a single solution of the HB equation for the MRFT to generate a stable periodic motion. For the system model presented in Eq. \eqref{eq_attitude_model_topdt} we have:
\begin{equation}
   \frac{\partial \Arg{G_{inner}(j\Omega)}}{\partial \Omega}=-\tau-\frac{T_{prop}}{1+(T_{prop}\Omega)^2}-\frac{T_{1}}{1+(T_{1}\Omega)^2}
\end{equation}
which is always negative, and hence satisfy the stability condition in Eq. \eqref{eq_loeb_criterion}. For the outer loop case, we need to analyze the closed loop dynamics given by:
\begin{equation}
    G_{cl}=\frac{C_{att}G_{att}}{1+C_{att}G_{att}}
\end{equation}
where \(C_{att}\) is a PD controller. Define \(G_{ol}=C_{att}G_{att}\). This closed-loop structure is detailed in Fig. \ref{fig:inner_loop_model} and shown in the context of the outer-loop in Fig. \ref{fig:outer_loop_model}. Given the properties of logarithmic derivatives of complex numbers, we have:
\begin{equation}
    \frac{1}{G}\frac{dG}{d\Omega}=\frac{d\ln{G}}{d\Omega}=\frac{d\ln{|G|}}{d\Omega}+j\frac{d\Arg{G}}{d\Omega}
\end{equation}
which leads to:
\begin{multline}
\label{eq_ho_aos_condition}
    \frac{d\Arg{G_{cl}}}{d\Omega}=\operatorname{Im}\{\frac{1}{G_{cl}}\frac{dG_{cl}}{d\Omega}\}\\
    =\operatorname{Im}\{\frac{1+G_{ol}}{G_{ol}}\frac{\frac{dG_{ol}}{d\Omega}(1+G_{ol})-\frac{dG_{ol}}{d\Omega}G_{ol}}{(1+G_{ol})^2}\}\\
    =\operatorname{Im}\{\frac{1}{G_{ol}(1+G_{ol})}\frac{dG_{ol}}{d\Omega}\}=\operatorname{Im}\{\frac{1}{1+G_{ol}}\frac{d\ln{G_{ol}}}{d\Omega}\} \\
    =\operatorname{Im}\{\frac{d\ln{G_{ol}}}{d\Omega}\}\operatorname{Re}\{\frac{1}{1+G_{ol}}\}+\operatorname{Re}\{\frac{d\ln{G_{ol}}}{d\Omega}\}\operatorname{Im}\{\frac{1}{1+G_{ol}}\} \\
    =\frac{d\Arg{G_{ol}}}{d\Omega}\operatorname{Re}\{\frac{1}{1+G_{ol}}\}+\frac{d\ln{|G_{ol}|}}{d\Omega}\operatorname{Im}\{\frac{1}{1+G_{ol}}\}
\end{multline}
This expression needs to be negative in order for the stability condition in Eq. \eqref{eq_loeb_criterion} to be satisfied. The terms \(\frac{d\Arg{G_{ol}}}{d\Omega}\) and \(\frac{d\ln{|G_{ol}|}}{d\Omega}\) in Eq. \eqref{eq_ho_aos_condition} are always negative as shown earlier. Because \(G_{cl}\) is always tuned for stability the gain margin would be greater than one. As a result the term \(\operatorname{Re}\{\frac{1}{1+G_{ol}}\}\geq 1\) is always positive for test parameter \(\beta\) close to zero. The term \(\operatorname{Im}\{\frac{1}{1+G_{ol}}\}\) is close to zero for the considered values of \(\beta\) which makes the first term in Eq. \eqref{eq_ho_aos_condition} dominant. As a result the expression in Eq. \eqref{eq_ho_aos_condition} is always negative which leads to the satisfaction of the asymptotic orbital stability condition given in Eq. \eqref{eq_loeb_criterion}. This finding was confirmed numerically for the range of parameters given in Table \ref{tab:parameters_range}.

%Ayyad notes:
\begin{comment}
Main ideas:
1- The probem is divided to two parts; first inner loop then outer loop.
2- Recap of finding distinguishing phase for inner loop systems
3- Assuming an inner loop system, a disnghuishing phase is obtained for the outer loop parameters.
\end{comment}

\section{Generating Representative Processes}
\label{sec:class_outputs}

%why
This section describes the steps undertaken to discretize the model parameter subspace \(D\) into a discretized set of representative processes \(\bar{D} := \{G_{1}, G_{2}, ..., G_{N}\}\) that capture the main dynamics of the full range of parameters shown in Table \ref{tab:parameters_range}. The discretization of \(D\) enables tackling parametric identification as a classification problem, where a classifier maps a process under test to the most appropriate process in \(\bar{D}\). The recognition of representative processes serves several objectives in the DNN-MRFT approach. First, it alleviates the need of an online real-time controller optimizer as optimal controllers are designed offline for all processes in \(\bar{D}\). Second, knowledge of the dynamics of \(\bar{D}\) is exploited to identify exact process gains as later explained in section \ref{sec:gain_correction}. The third objective of discretizing \(D\) is providing a measure of discrepancy between all representative processes that correspond to the controller auto-tuning objective. This measure of discrepancy is utilized for the training of DNN-classifiers as discussed in detail in section \ref{sec:data_gen}. These advantages outweigh the marginal loss of accuracy resulting from the discretization process, which is shown to be negligible by the results obtained in Table \ref{tab:simulation_results}.

%The discritization of \(D\) serves two purposes in the DNN-MRFT approach. First, it enables tackling parametric identification as a classification problem where the representative processes become the contending outputs of a classifier. Despite the recent developments in DL for regression problems, DL is still mostly adopted in classification applications due to .... As such, the parametric identification problem becomes that of mapping a process under test to its best representative process \(G_{output} \in \bar{D}\). Secondly, optimal controllers can be designed offline for each member process in \(\bar{D}\) forming a lookup table of optimal controller parameters \(\bar{C^{*}}\). As a result, near-optimal controllers with well-defined performance and stability margins can instantaneously be identified during the DNN-MRFT adaption phase without the need of an online optimization process.

%discritization criterea
A proper criteria must be defined for the discretization of \(D\) that establishes sufficient guarantees on the performance of system identification and controller auto-tuning without sacrificing the distinguishability of the discretized processes. For instance, an equispaced discretization with a small partitioning distance would generate an overly-discretized \(\bar{D}\) with an imbalanced representation of the frequency response characteristics of \(D\); this in turn introduces undesirable biases in training the DNN classifiers. Alternatively, a very large partitioning distance generates substantial discretization errors and does not guarantee proper performance margins for the optimal controllers designed offline for \(\bar{D}\). As our objective is auto-tuning controller parameters, for the criterion of discretization, we adopt the concept of controller performance deterioration used for the system identification approach presented in \cite{ayyad2020real}. Given a performance index \(Q\) that quantifies errors resulting from a closed loop application of controller \(C\) to a process \(G\), the controller performance deterioration \(J_{ij}\) between two dynamic processes \(\{G_i,G_j\}\) is defined as:

\begin{equation}\label{eq_performance_deterioration}
J_{ij} = \frac{Q(C_i^{*}, G_j) - Q(C_j^{*}, G_j)}{Q(C_j^{*}, G_j)} \times 100 \%
\end{equation}
where \(J_{ij}\) represents the relative degradation in performance in terms of \(Q_{ISE}(C_{i}^{*}, G_i)\) when the optimal controller of \(G_{i}\) is replaced by \(C_{j}\), which is optimal for another process \(G_{j}\). Eq. \eqref{eq_performance_deterioration} is the relative sensitivity function introduced in \cite{rohrer1965sensitivity}. It must be noted that the above formulation of the controller deterioration is non-commutative, that is \(J_{ij}\neq J_{ji}\). Therefore, the joint cost function \({J}^{max}_{(ij)}=max\{J_{ij}, J_{ji}\}\) is used as the discretization criteria in the remainder of this paper. Additionally, the design of optimal controllers is limited to a PD structure with a minimum phase margin constraint imposed to the controller optimization problem. The performance index \(Q\) used for controller synthesis is the conventional ISE criterion applied to a unit step response, and is given by:

\begin{equation}\label{eq_ise}
Q_{ISE}(C, G) = \frac{1}{T_s}\int_{0}^{T_s} e(t)^2 dt
\end{equation}

\begin{figure}[t]
\begin{center}
\includegraphics[width=\linewidth]{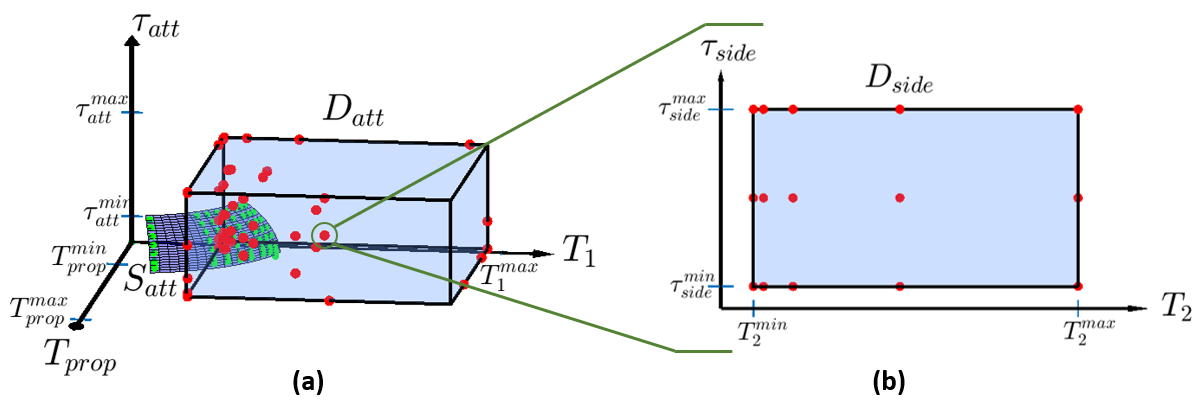} 
\end{center}
\caption{The overall discretization scheme for identifying key processes in the parameter space. (a) Inner loop dynamics are first discretized into \(\bar{D}_{att}\) according to the principle of controller performance deterioration. (b) For each member process of \(\bar{D}_{att}\), a different set of discretized outer loop model parameters is identified. The output of the process would be \(N_{att}\) sets of outer loop discrete processes.}
\label{fig:full_discritization_process}
\end{figure}

Following the criterion in Eq. \eqref{eq_performance_deterioration}, discretization is performed such that adjacent processes in \(\bar{D}\) achieve a target joint cost \(J^*\) within an admissible tolerance value. We first discretize the three-dimensional parameters space of altitude and attitude models. For computational efficiency we have utilized the linear time scale property of cost function \(Q(C(T_cs),G(T_cs))=\alpha_cQ(C(s),G(s))\), where \(\alpha_c\) is an unknown dimensionless constant, to perform sampling in a two-dimensional hemispherical hyper-surface \(S\) as shown in part (a) of Fig. \ref{fig:full_discritization_process}. Those discretized processes in \(S\) get scaled in time to populate \(\bar{D}\) as illustrated in Fig. \ref{fig:full_discritization_process} and detailed in our previous work \cite{ayyad2020real}.Based on the model parameter bounds given in Table \ref{tab:parameters_range} and the discretization accuracy specifications given in Table \ref{tab:discritization_specs}, discretization of \(D_{alt}\) and \(D_{att}\) yield a total number of \(N_{alt}=208\) and \(N_{att}=48\) representative processes respectively.

The discretization of \(D_{side}\) depends on the identified inner loop process and its corresponding optimal controller. Hence \(\bar{D}_{side,i}\) is unique for every process \(G_{att,i} \in \bar{D}_{att}\) where \(i \in {1,...,N_{att}}\). The discretization of a given \(D_{side}\) domain is performed based on the maximum sensitivity values obtained from \(\frac{\partial J_{ij}}{\partial T_2},\,\frac{\partial J_{ij}}{\partial \tau_{side}}\). Spacing of the discretization was based on the most sensitive \(J_{ij}\) to changes in the parameters space which resulted in a slightly over-discretized \(\bar{D}_{side}\) but guarantees that the cost between adjacent processes does not exceed \(J^*\).

To summarize, discretization steps can be stated as follows:
\begin{enumerate}
    \item Identify domain of discretization \(D_{alt}\), \(D_{att}\), and \(D_{side}\).
    \item Set the three relative sensitivity parameters: \(J^*\), admissible tolerance, and minimum phase margin \(\phi_m\) as shown in Table \ref{tab:discritization_specs}.
    \item Discretize \(D_{att}\) and \(D_{alt}\) as illustrated in Fig. \ref{fig:full_discritization_process} and outlined in \cite{ayyad2020real}.
    \item Run Algorithm \ref{alg:outer_loop_disc} on \(D_{att}\) to get \(\bar{D}_{side}\) and \(\bar{C}^*_{side}\).
    
\end{enumerate}

\renewcommand{\arraystretch}{1.5}
\begin{table}[]
    \caption{Specifications for the process of discretizing the parameter space \(D\)}
    \centering
    \begin{tabular}{|p{3.5cm}|c|}
        \hline
        Target joint cost \(J^{*}\) & \hspace{1.7cm} 10\% \hspace{1.7cm} \\
        \hline
        Admissible tolerance  & 3\% \\
        \hline
        Minimum \(PM\) constraint & 20 \\
        \hline
        Optimization algorithm for controller design & Nelder-Mean simplex algorithm\\
        \hline
    \end{tabular}
    \label{tab:discritization_specs}
\end{table}

\begin{algorithm}[h]
\SetAlgoLined
\begin{flushleft}
        \textbf{INPUT:} \((\bar{D}_{att},D_{side}, J^{*})\) - Attitude loop set of discretized parameters, parameters domain of outer loop dynamics, target joint cost \\
        \textbf{OUTPUT:}  \((\bar{D}_{side},\bar{C}_{side}^{*})\) - Set of outer loop discretized processes, Lookup table of outer loop optimal controller parameters
\end{flushleft}
\begin{algorithmic}[1]
\FORALL{\(G_{att,i} \in \bar{D}_{att}\)}
\STATE Identify inner loop optimal controller \(C_{att,i}^{*}\)\;
\STATE Utilizing \({C_{att,i}^{*}, G_{att,i}}\), discretize \(D_{side}\) into \(\bar{D}_{side,i}\) based on \(J^*\)\;
\STATE \(\varphi_{d,i} \gets GetDistinguishingPhase({C_{att,i}^{*}, G_{att,i}},D_{side})\) from Algorithm \ref{alg:dist_phase}
\FORALL{\(G_{side,ij} \in \bar{D}_{side,i}\)}
\STATE Identify outer loop optimal controller \(C_{side,ij}^{*}\)\;
\STATE \(C_{side,i}^{*} \gets C_{side,i}^{*} \cup C_{side,ij}^{*}\)\;
\ENDFOR
\STATE \(\bar{D}_{side} \gets \bar{D}_{side} \cup \bar{D}_{side,i}\)\;
\STATE \(C_{side}^{*} \gets C_{side}^{*} \cup C_{side,i}^{*}\)\;
\ENDFOR
\end{algorithmic}
\caption{Side motion parameters domain discretization}
\label{alg:outer_loop_disc}
\end{algorithm}

\section{Data Generation and Deep Neural Network Model Training}
\label{sec:data_gen}

%objective
The deep neural network component of the DNN-MRFT approach provides a mapping from the MRFT response of the unknown process to the best representative process in \(\bar{D}\). This mapping is denoted by \(\Gamma: X \rightarrow \bar{D}\); where \(X \in R^{2\times n_{s}}\) is a vector concatenating \(n_{s}\) samples of the controller output and process variable of the MRFT response. We have previously demonstrated the appropriateness of DNN for the system identification task in \cite{ayyad2020real}, where a single network was utilized for the identification of attitude and altitude model parameters. In this section, we build upon our previous results and present a multi-network solution for the full identification of UAV dynamics. 

%A different network for each loop.
The classification outputs generated in Section \ref{sec:class_outputs} fall into three sets of model parameters, which requires three different mappings to be solved. We train a unique DNN classifier for each of these mappings. One challenge however is the dependency of the outer loop system response on the inner loop dynamics, which results in multiple variations of \(\bar{D}_{side}\) as demonstrated in Section \ref{sec:class_outputs}. Similarly, different inner-loop processes would result in a different distinguishing phase for the outer-loop model parameters, which in turn alters the criteria for generating the DNN input vector \(X\). Changes in the classifier's input and output layer due to the inter-loop dependencies make the utilization of a single DNN network for the identification of side-motion model parameters impractical. Rather, we employ \(N_{att}=48\) DNN networks for the outer-loop identification problem, each assuming a specific inner-loop process \(G_{att}\in\bar{D}_{att}\). In total, 50 DNNs are trained: one for altitude dynamics, one for attitude dynamics, and 48 for side-motion dynamics.

\begin{figure}[t]
\includegraphics[width=\linewidth]{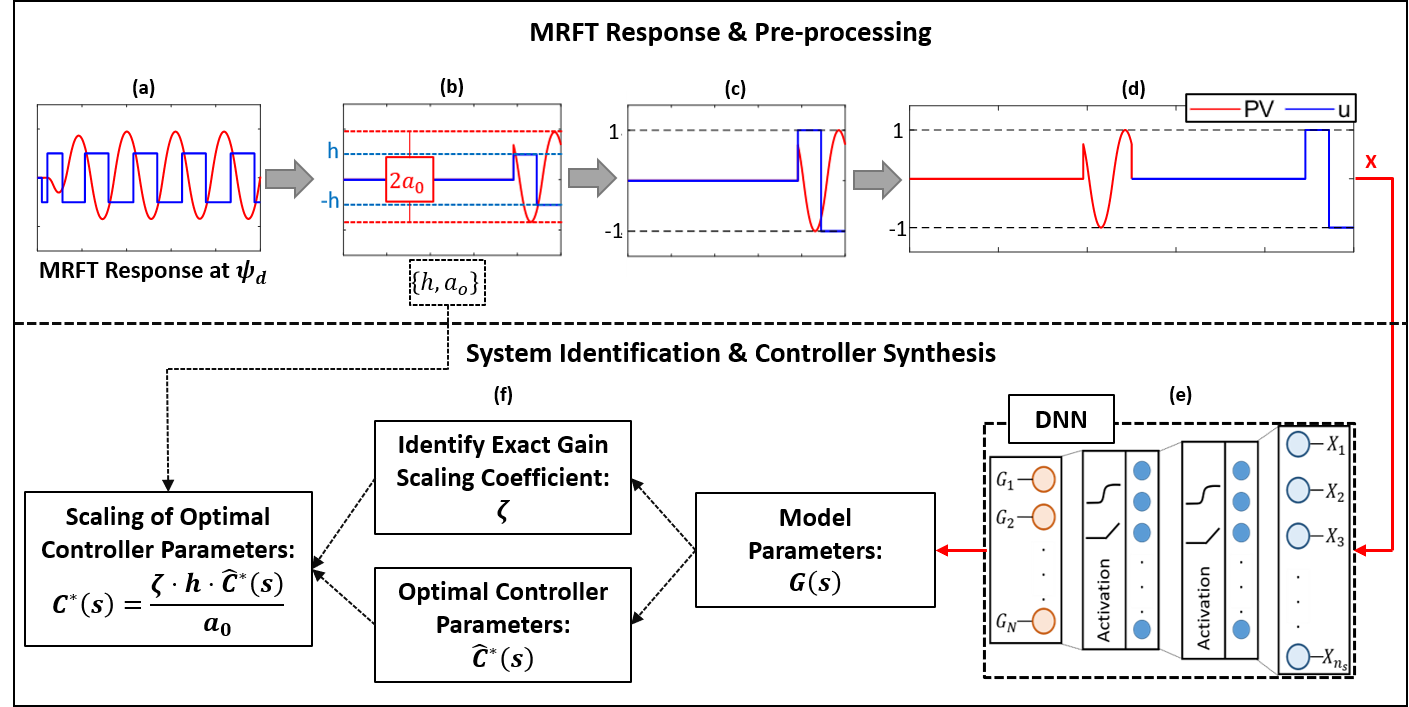} 
\caption{The overall DNN-MRFT pipeline showing the steps of obtaining and pre-processing the MRFT response followed by system identification and controller synthesis. (a) The process's MRFT response is obtained at the distinguishing phase and the sampling time is adjusted to be 1ms. (b) One cycle of the steady-state oscillation is selected, zero-padding is applied elsewhere. (c) The response is zero-centered and scaled to a unity amplitude. (d) \(PV\) and \(u\) are concatenated to form the DNN input vector \(X\). (e) The DNN network corresponding to the proper control loop is selected and used to predict the model parameters \(\hat{G}(s)\). The DNN structure consists of a sequence of fully-connected layers and activation functions. (f) From a lookup table, the gain-normalized optimal controller parameters and the exact gain scaling coefficient \(\zeta\) are found. \(\zeta, h\) and \(a_{0}\) are then used to properly scale the controller parameters. }
\label{fig:dnn_preprocess}
\end{figure}

%Training data & pre-processing
Training data for the classification problem was generated in simulation for all member processes in \(\bar{D}\). For each \(G_{i} \in \bar{D}\), the MRFT response with parameter \(\beta\) set to the corresponding distinguishing phase was simulated 30 times with randomly varied measurement noise \(\aleph\) and input biases \(u_{0}\) to generate the DNN training set. The incorporation of imperfections like \(u_{0}\) and \(\aleph\) prompts regularization and generalization to varied experimental conditions during the training process \cite{Noh2017}. The maximum value of \(u_{0}\) was constrained to half the relay amplitude \(h\) of the MRFT controller as a reasonable bias magnitude in practical settings. A validation set was also generated in a similar manner for hyper-parameter tuning and evaluation purposes. The validation set consist of 15 simulations per candidate process. The DNN input vector \(X\) is obtained by processing the MRFT response according to the following steps: sampling adjustment, cropping, zero-padding, amplitude normalization, and concatenation. The size of the input vector \(X\) is determined by the slowest MRFT response within the corresponding parameter set \(\bar{D}\). Fig. \ref{fig:dnn_preprocess} illustrates our overall pipeline of UAV system identification and controller tuning using deep neural network. 

\begin{comment}

\begin{figure}[t]
\includegraphics[width=\linewidth]{NNstructure.png} 
\caption{DNN model architecture consisting of a sequence of fully-connected layers and activation functions.}
\label{fig:DNN_structure}
\end{figure} 
\end{comment}

%Network training (structure, hyper-parameters, and cost function) Sweeping for parameters
All the developed DNN models follow the same architecture shown in Fig. 
\ref{fig:dnn_preprocess}, where sequences of fully-connected layers and activation functions are concatenated. Dropout and batch normalization are applied to the outputs of each fully-connected layer to avoid over-fitting and accelerate the training process \cite{Li2017, Ioffe2015}. After the final fully-connected layer, we utilize the cost-augmented soft-max formulation introduced in \cite{ayyad2020real}, which exhibited performance improvements over the conventional soft-max formulation for system identification tasks due to introducing meaningful discrepancies to the cost of miss-classification. The augmented formulation is given by: 

\begin{equation} \label{eq_modified_softmax}
p_{i} = \frac{e^{(1+J_{iT})\cdot a_{i}}}{ \sum_{j=1}^{N}  e^{(1+J_{iT})\cdot a_{j}}}
\end{equation} 
where the controller deterioration joint cost \(J_{iT}\) is utilized as the measure of discrepancy between the DNN prediction and the ground truth model parameters \(G_{T}\). Cross-entropy is then utilized as the loss function for training the DNN models.

We utilized the ADAM optimization algorithm for training as it is a well-established algorithm with proven advantages in terms of convergence speeds and robustness to noisy gradients \cite{Shrestha2019, kingma2014}. We implemented an automated search approach to determine the best network size and set of hyper-parameters for each of the developed 50 DNN models. The variables included in the search process along with their corresponding search space are shown in Table \ref{tab:dl_search}. For each classification task, the network that performed best on the validation set was selected as the preeminent DNN model.

\renewcommand{\arraystretch}{1.5}
\begin{table}[h!]
    \caption{Search space of DNN structure and hyper-parameter optimization process}
    \centering
    \begin{tabular}{|p{0.4\linewidth}|m{0.4\linewidth}|}
        \hline
        \textbf{Parameter} & \textbf{Search Space} \\
        \hline
        Number of layers & \{1, 2, 3\} \\
        \hline
        Neurons per layer & \{50, 100, 1000, 3000\} \\
        \hline
        Activation function & \{ReLU, tanh\} \\
        \hline
        Base learning rate & 0.005 \\
        \hline
        Gradient decay factor & 0.9 \\
        \hline
        Gradient decay factor & 0.999 \\
        \hline
    \end{tabular}
    \label{tab:dl_search}
\end{table}

%Performance on validation set?? Average and worst case?

\section{Identification of Exact Process Gain}
\label{sec:gain_correction}
The information contained in the MRFT response of the system are embedded within the generated oscillations in three forms: frequency, amplitude, and shape. The DNN classifier only utilizes the frequency and shape of the oscillation to identify model parameters as the amplitude is normalized to one during pre-processing. The importance of the shape of the oscillation emphasizes the fact that the MRFT excites multiple frequencies of the linear system. The periodic components of the self-excited oscillation can be given by the Fourier series \cite{atherton1975}:
\begin{equation}
    y(f)=\sum_{n=0}^{\infty} a_n cos(nf)+b_n sin(nf)
\end{equation}
where \(a_n\) and \(b_n\) are Fourier series coefficients. For an odd symmetric nonlinearity (note that MRFT switching at steady state resembles a hysteresis relay), coefficients for even values of \(n\) and other odd harmonics exist. The DF solution, presented in Eq. \eqref{eq_mrft_df} for the MRFT, accounts for the first order harmonic only. The amplitude of the harmonics depend on the low pass filtering properties for every process in \(\bar{D}\). Therefore, if we have identified process parameters experimentally, we can use such knowledge to predict exact system amplitude response. Exact analytical solution of Lure systems can be provided by the LPRS method \cite{boiko1999lprs,boiko2008discontinuous}, or Tsypkin's method \cite{boiko2007chattering}. To achieve exactness and real-time capability, we simply introduce a scaling coefficient \(\zeta\) that provides exact system gain for every system in \(\bar{D}\) and make these values available in a look-up table. The values of \(\zeta\) are found by simulating MRFT control with each system in \(\bar{D}\). In simulation, we use the same MRFT implementation used experimentally and measure the system steady-state response amplitude to find \(\zeta\). During the DNN-MRFT identification phase, the proper \(\zeta\) value is selected from the look-up table and is used to scale the identified controller parameters as shown in Fig. \ref{fig:dnn_preprocess}. The results in Table \ref{tab:simulation_results} show the improvement in controller performance on a simulated test set due to the use of identified exact gain scale compared to amplitude reported by the DF method. Note that the DF method uses \(\zeta=\frac{4}{\pi}\) for all processes.

\section{Take-off Controller Design }
\label{sec:take_off_cont}
Using the trained DNN and MRFT, the UAV can takeoff and immediately identify UAV dynamic parameters. MRFT parameter \(\beta\) corresponds to the distinguishing phase and was found in Section \ref{sec:dist_phase}. MRFT parameter \(h\) needs to be designed such that it provides adequate amplitude response and robustness against sensor noise and model biases. When MRFT is running, we can consider the following simplified system structure:
\begin{equation}
    \dot{x}=Ax(t)+B(u+f_d)
\end{equation}
where \(|f_d|\leq L>0\) is unknown bounded constant bias inherent in the system. Periodic oscillations would be achieved whenever \(u_0-h\leq f_d \leq u_0+h\) where biased MRFT output is given by \(u_M=u_0 \pm h\). In this section we suggest algorithms to ensure: (1) condition \(u_0-h\leq f_d \leq u_0+h\) is met while (2) avoiding excessive \(h\) values that would result in impractical amplitudes of the oscillations like roll angles beyond \(\frac{\pi}{2}\). The procedure for the design of these algorithms can cover any size and configuration of multirotor UAVs.
Bias caused by the gravity makes identification of altitude dynamics particularly challenging. The elimination of the gravity bias without prior knowledge of a UAV's total generated thrust and mass requires an algorithm that can handle the take-off state. For that we use a cascaded switched PID controller as shown in Fig. \ref{fig:identification_scheme} by adding switching position \(1^*\) to altitude. In the first stage (\(S\) is at \(1^*\)), a PI controller is used for take-off:
\begin{multline}\label{eq_takeoff_algorithm}
u_z(t)=
\left\{
\begin{array}[r]{l l}
K^{tf}_c e_z(t)+ \int_{0}^{T}K^{tf}_i e_z(t) dt : \Ddot{z_I}\leq g+\delta\\
U_M(e_{z}(t),h_{alt})+u_{z0}  : C
\end{array}
\right.
\end{multline}
where \(\delta>0\) is a bias factor to compensate for increased efficiency in take-off due to ground effect, \(u_{z0}\) is the last output of the PI controller, and \(C\) is a condition that is set permanently to true once the condition in the first line is violated. The first line of Eq. \eqref{eq_takeoff_algorithm} corresponds to \(1^*\) position of \(S\) in Fig. \ref{fig:identification_scheme}, while the second line corresponds to switch position \(1\). The switching condition aims at minimizing the value of the bias present in the MRFT switching, which is perfectly achieved when \(u_{z0}\) produces a thrust that causes the UAV to hover. It is not always possible to have a clean measurement of acceleration which was the case in our experimental setup, and therefore we were using position measurement. Though position measurement is lagged by a phase of \(\pi\), take-off can be slowed down and a value of \(u_{z0}\) close to hover thrust can still be achieved. For position measurement case, the condition in the first line of Eq. \eqref{eq_takeoff_algorithm} would be \(z_I \leq \delta_p\) instead of \(\Ddot{z_I}\leq g+\delta\).

Because the presented controllers in Eq. \eqref{eq_takeoff_algorithm} will be applied for all multirotor UAVs with the full model parameter's range in Table \ref{tab:parameters_range}, suitable values of the take-off controller parameters need to be designed. The optimization decision variables are the parameters \(K^{tf}_c\), \(K^{tf}_i\), \(\delta_p\), and \(h\) presented in Eq. \eqref{eq_takeoff_algorithm}. A cost function has been designed to address the optimization of these values:
\begin{align}
\begin{split}\label{eq_takeoff_cost}
J_{bias}=&\sqrt{\frac{(\frac{t_h}{t_h+t_l}-t_{b0})^2}{t_{b0}^2}}\\
J_{time}=&\begin{cases}
      0 & : t_r < t_{r0} \\
      \sqrt{\frac{(t_r-t_{r0})^2}{t_{r0}^2}} & : t_r \geq t_{r0}
    \end{cases}\\
J_{amp}=&\begin{cases}
      0 & : a_r < a_{r0} \\
      \sqrt{\frac{(a_r-a_{r0})^2}{a_{r0}^2}} & : a_r \geq a_{r0}
    \end{cases}\\
J_{tot}=&J_{bias}+J_{time}+J_{amp}
\end{split}
\end{align}
where \(J_{bias}\), \(J_{time}\), \(J_{amp}\) are the costs associated with bias in relay, system rise-time, and excited process amplitude respectively. \(t_{h}\) is the duration MRFT switches high, \(t_{l}\) is the duration MRFT switches low, and \(t_{b0}=0.5\) is a constant that corresponds to the case when MRFT switching is symmetric, i.e. \(t_{h}=t_l\rightarrow t_{b0}=\frac{t_{h}}{t_{h}+t_{l}}\). The value of \(t_r\) is the time it takes to reach \(90\%\) of desired altitude from take-off (take-off is defined as passing 2cm altitude), and \(t_{r0}\) corresponds to the desired maximum rise time which we chose to be 5s. The value of \(a_r\) corresponds to the steady-state amplitude of the self-excited oscillation due to MRFT. The value \(a_{r0}\) corresponds to the desired maximum MRFT amplitude response and was chosen to be 0.3m. The selection of cost function input parameters reflects essential practical requirements of the auto-tuner. We found that a severely biased relay might force motors to function near their operational extremes. A long take-off time is not desired and can be dangerous due to the fact that at take-off, MRFT is also running on roll and pitch where rotor tips might hit the floor. The last risk accounted for is associated with excessively large amplitudes of the response, which might lead to crashes or undesired aggressiveness. The collective responses of systems at the vertices of \(\bar{D}\) (actually \(D\) resembles a cuboid in the system parameters space) was used to find \(J_{tot}\). We found that this optimization problem is non-convex so that multiple initial points were tested. Nelder-Mead simplex algorithm realized by "fminsearch" function in MATLAB\textregistered \(\,\)has been used. The resulted optimal decision variables are:
\begin{multline}
        \label{eq_opt_takeoff_para}
    h=0.10746,\;K^{tf}_c=9.4969\times 10^{-2},\;\\
    K^{tf}_i=9.8754\times 10^{-3},\; \delta_p=0.11984
\end{multline}

The responses of systems at the vertices of \(\bar{D}\) to the take-off algorithm with optimal take-off parameters can be seen in Fig. \ref{fig:takeoff_opt} in the appendix, where it can be clearly seen that all UAV variants are stable and operating within the physical limits. Note that some systems take very long to start taking-off compared to \(t_{r0}\). This is due to the use of the position measurement for \(\delta_p\) instead of acceleration measurement. Though the values presented in Eq. \eqref{eq_opt_takeoff_para} guarantee stability, tuning from take-off can be made faster and smoother with smaller amplitude of the excited oscillations and faster take-off time. We find this possible with prior knowledge of the peak-thrust to weight ratio \(C_{TW}\) of the UAV (the considered range of \(C_{TW}\) based on Table \ref{tab:parameters_range} is 1.5 to 5). The value of \(C_{TW}\) is easy to find (i.e. motor datasheet and a weighing scale) which makes the suggested auto-tuner still suitable for non-experts. The optimization of the take-off parameters can be run again while using systems in \(\bar{D}\) which satisfies selected \(C_{TW}\) value.
\begin{comment}
Due to inherent system biases (e.g. weight imbalance, sensor miscalibration, etc.) and due to the fact that the take-off algorithm can get affected by disturbances external to the system (e.g. ground effect, wind, etc.), the amplitude and bias of the system response might become excessive according to the criteria presented in Eq. \eqref{eq_takeoff_cost}. Because this might affect identification accuracy, we designed a simple algorithm that succeeds the take-off algorithm and adjusts MRFT amplitude and bias based on previously excited stable oscillations.
\end{comment}
Other control loops have smaller bias values and MRFT can be started immediately. To design suitable \(h\) value for roll and pitch control loops we consider the parameters ranges in Table \ref{tab:parameters_range} and a limit of \(30^\degree\) on the maximum expected oscillation's amplitude. A suitable selected \(h\) value for roll and pitch is \(5\%\).

The UAV is prone to drift side ways as the outer-loop control is not active during roll and pitch identification. This is not an issue when plenty of space is available, e.g. when the identification is performed in an open space. But when identification needs to be performed in a constrained space, a temporary outer loop controller is needed to bound the UAV to the vicinity of the take-off location until roll and pitch identification is completed. We suggest the following \textit{bounding} controller composed of a hysteresis relay cascaded with a relay with dead-band:
\begin{multline}\label{eq_bounding_box}
u_{bb}(t)=\\
\left\{
\begin{array}[r]{l l}
0\; &:\, |e(t)| \leq \varepsilon_1 \; \land\; u_{bb}(t-) = 0 \\
h_{o1} sgn(e(t)) \; &:\, |e(t)| \geq \varepsilon_1 \;\land\; u_{bb}(t-) = 0\\
h_{o1} sgn(e(t)) \; &:\,|e(t)| \leq \varepsilon_2 \;\land\; u_{bb}(t-) \neq 0\\
(h_{o1}+h_{o2}) sgn(e(t)) &:\, |e(t)| \geq \varepsilon_2\\
\end{array}
\right.
\end{multline}
where \(\{\varepsilon_1,\varepsilon_2\}\) are predefined position thresholds, and \(h_{o1}\) and \(h_{o2}\) are the amplitudes of the hysteresis relay and the relay with dead-band respectively. To guarantee that inner-loop MRFT oscillations reach steady-state, the threshold on position must be large enough such that the switching frequency of the bounding controller is considerably lower than that of the inner-loop oscillations. Table \ref{tab:takeoff_parameters} lists all parameters specific to the take-off stage that can be used across any multirotor UAV design and size within the ranges considered in Table \ref{tab:parameters_range}

\renewcommand{\arraystretch}{1.5}
\begin{table}[h!]
    \caption{Summary of take-off parameters}
    \centering
    \begin{tabular}{|p{0.06\textwidth}|p{0.105\textwidth}|p{0.24\textwidth}|}
        \hline
        Parameter & Value (Unit) & Comment \\
        \hline
        \(h_{alt}\) & \(10.746\) (\% ESC)  & See Eq. \eqref{eq_opt_takeoff_para}\\
        \hline
        \(K^{tf}_c\) & \(9.4969\times 10^{-2}\)  & See Eq. \eqref{eq_opt_takeoff_para} \\
        \hline
        \(K^{tf}_i\) & \(9.8754\times 10^{-3}\)  & See Eq. \eqref{eq_opt_takeoff_para}\\
        \hline
        \(\delta_p\) & \(0.11984\) & See Eq. \eqref{eq_opt_takeoff_para} \\
        \hline
        \(h_{att}\) & \(5.0\) (\% ESC) & Same for roll and pitch \\
        \hline
        \(h_{o1},h_{o2}\) & \(0.05\) (rad) & Same for roll and pitch \\
        \hline
        \(\epsilon_{1},\epsilon_{2}\) & \(1,2\) (m) & Selection depends on space constraints  \\
        \hline
    \end{tabular}
    \label{tab:takeoff_parameters}
\end{table}

\section{Simulation and Experimental Results}

This section presents the simulation and experimental evaluation of the DNN-MRFT approach. Both evaluation methods follow the protocol demonstrated in Fig. \ref{fig:identification_scheme}. DNN-MRFT is first used to identify altitude and attitude model parameters \(\{\hat{G}_{alt}, \hat{G}_{att}\}\) and their corresponding optimal controllers \(\{\hat{C}^{*}_{alt}, \hat{C}^{*}_{att}\}\). Then depending on the estimated inner-loop dynamics, outer-loop distinguishing phase and DNN classifiers are selected and applied immediately to side-motion dynamics to identify \(\hat{G}_{alt}\) parameters and optimal controller \(\hat{C}^{*}_{side}\) gains. The analysis presented in this section assesses the DNN-MRFT approach for: accuracy and persistence of the parametric identification,  the capability of the take-off controller to successfully lift and stabilize a UAV with no prior knowledge of system dynamics, and the performance of the auto-tuned controllers in aggressive trajectory following maneuvers.

Experimental tests were conducted on a variety of UAV designs as shown in Table. \ref{tab:uav_specs}. For clarity, we only present results obtained on the QDrone in this section while the remaining experimental results can be found in Fig. \ref{fig:exp_tuning_full_djif550} in the appendix and are also presented in the companion video \cite{paper_video}. Optitrack's motion capture system was used for UAVs localization \cite{Optitrack}. 

\renewcommand{\arraystretch}{1.5}
\begin{table}[h!]
    \caption{Specifications of the UAV designs used in experimental analysis}
    \centering
    \begin{tabular}{|p{1.6cm}|p{1.8cm}|p{1.8cm}|p{1.8cm}|}
        \hline
        & QDrone \cite{qdrone} & DJI F550 & DJI F550 with extended arms \\
        \hline
        Dimensions ($cm$) & \(40 \times 40 \times 15\)  & \(79 \times 72 \times 27\) & \(111 \times 100 \times 27\)\\
        \hline
        Mass ($kg$) & 1.0  & 2.09 & 3.38 \\
        \hline
        $\{I_{x},I_{y}, I_{z}\}$ & $\{0.010,$ $0.008, 0.015\}$ & $\{0.031,$ $0.030, 0.052\}$ & $\{0.093,$ $0.089, 0.156\}$\\
        \hline
        Number of propellers & 4 & 6 & 6 \\
        \hline
        Processor & Intel Aero Compute Board & Raspberry Pi 3 B+ & Raspberry Pi 3 B+ \\
        \hline
    \end{tabular}
    \label{tab:uav_specs}
\end{table}

\subsection{Persistence and Accuracy of Identification}

\subsubsection{Simulation Results}
\label{subsubsec:sim_results}
\renewcommand{\arraystretch}{2}
\begin{table*}[h]
\caption{Simulation results of the DNN-MRFT approach on 500 randomly selected processes in \(D\)}
    \centering
 \begin{tabular}{|p{3.5cm}|>{\centering\arraybackslash}m{1.2cm}|>{\centering\arraybackslash}m{1.2cm}|>{\centering\arraybackslash}m{1.2cm}|>{\centering\arraybackslash}m{1.2cm}|>{\centering\arraybackslash}m{1.2cm}|>{\centering\arraybackslash}m{1.2cm}|>{\centering\arraybackslash}m{1.2cm}|>{\centering\arraybackslash}m{1.2cm}|} 
 \hline
  \backslashbox{Gain scaling}{Criterion} & Average \(J_{att}\) & Maximum \(J_{att}\) & Average \(PM_{att}\) & Minimum \(PM_{att}\) & Average \(J_{side}\) & Maximum \(J_{side}\) & Average \(PM_{side}\) & Average Minimum \(PM_{side}\) \\
 \hline
 Normalized gain & 0.45\% & 5.10\% & 19.65 & 14.21 & -0.19\% & 4.91\%  & 19.70 & 17.18 \\
 \hline
 DF gain approximation & 1.45\% & 9.68\% & 20.04 & 14.20 & 2.04\% & 57.44\%  & 19.46 & 16.11 \\
 \hline
 Exact gain scaling & -1.77\% & 7.14\% & 18.41 & 11.24 & -1.73\% & 15.68\%  & 18.41 & 14.23 \\
 \hline
\end{tabular}
\label{tab:simulation_results}
\end{table*}

The objective of simulation analysis is to evaluate the optimality of the DNN-MRFT auto-tuned controllers for the full parameters' range presented in Table \ref{tab:parameters_range} with exact knowledge of the ground truth model. Five hundred different model parameter sets were randomly sampled from \(D\) to form a testing set \(\bar{D}_{test}\). For each \(G_{T}\in \bar{D}_{test}\), the DNN-MRFT approach predicts a process \(G_{p}\) and a controller \(C^{*}_{p}\) for both the inner and outer control loops under randomly varied conditions of noise \(\aleph\) and bias \(u_0\). We utilize the controller deterioration criterion \(J_{pT}\) from Eq. \eqref{eq_performance_deterioration} to quantify the accuracy of \(G_{p}\) estimation. Additionally, the phase margin \(PM\) of \(C^{*}_{p}(s) G_{T}(s)\) is presented to assess the robustness of the synthesized controller. Average and worst-case results on the entire testing set are reported in Table \ref{tab:simulation_results}. The worst-case phase margin of the side-motion control loop is reported as the average of the worst-case prediction of each of the 48 outer-loop DNNs. Results are reported with two different gain scaling methods: the first method uses the DF approximation with \(\zeta=\frac{4}{pi}\) to approximate the gain of the unknown system, and the second one uses the exact scaling method described in Section \ref{sec:gain_correction}. In Table \ref{tab:simulation_results}, the case when the gain is normalized (does not include errors introduced by gain scaling) is also presented for comparison.

The results in Table \ref{tab:simulation_results} demonstrate two features of DNN-MRFT. First it demonstrates stability by the guarantee of a minimum phase margin value, which agrees with the original formulation of the MRFT and coordinated test and tuning \cite{Boiko2013}. This is clearly indicated by the minimum phase margin values that are always positive. Note that the average \(J_{att}\) and \(J_{side}\) values when exact gain scaling is used are negative. Negative joint cost values occur whenever tuning resulted in a phase margin that is lower than the minimum phase margin constraint used in parameter space discretization process and shown in Table \ref{tab:discritization_specs}. The second feature of DNN-MRFT is that it results in near-optimal performance for the full range of \(D\). Average controller deterioration cost for both inner-loop and outer-loop dynamics are near zero, with the worst-case deterioration being \(15.68\%\) for the side-motion auto-tuning case. Furthermore, our proposed DF gain scaling approach results in a generally lower deterioration than the DF approximation, especially when considering worst-case results. These results show an inherent advantage of using DNN-MRFT: it employs nonlinear methods (i.e. MRFT and DNN) to obtain a linear description of the underlying dynamics, which preserves the useful properties of linear systems such as the measurable robustness and performance margins. 

\subsubsection{Experimental Results} \label{sec:y_ID}

\begin{figure}[h]
\includegraphics[width=\linewidth, trim={0cm 0cm 0cm 0cm}]{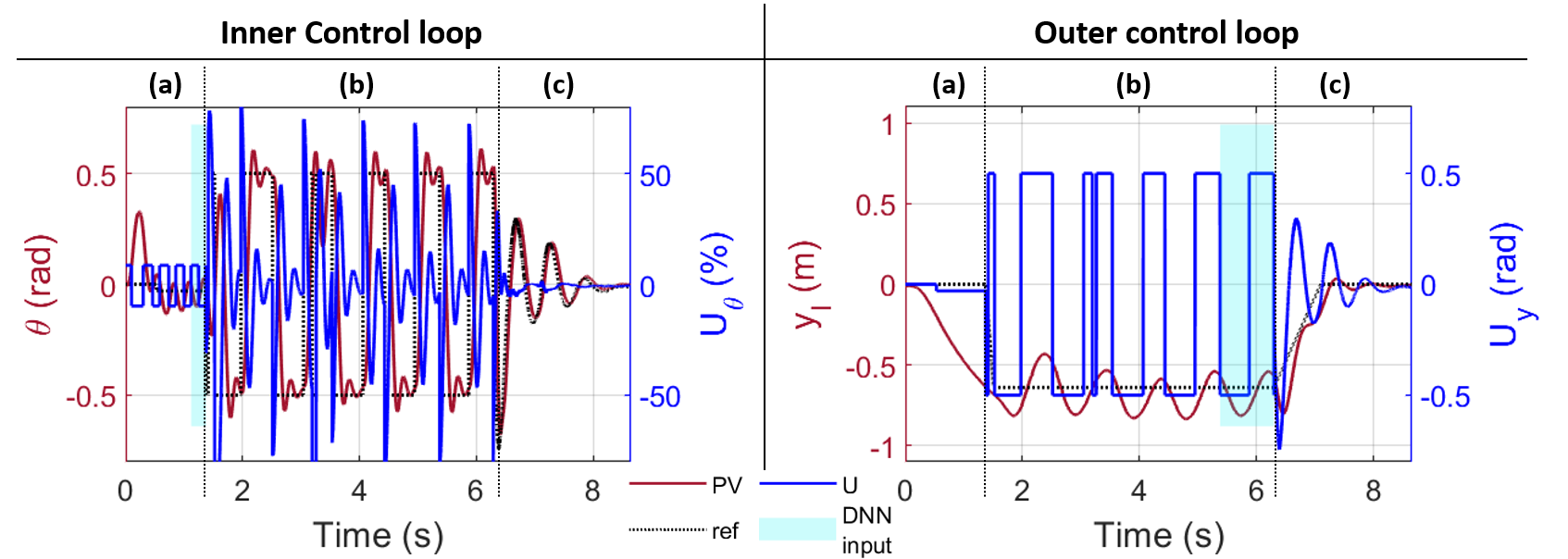} 
\caption{DNN-MRFT auto-tuning experiment for the inner and outer control loops. (a) MRFT is performed on the inner-loop until steady-state oscillations are acquired. The last cycle is passed to a DNN that predicts model and controller parameters. (b) DNN-MRFT is repeated for the identification and tuning of side motion dynamics. (c) The UAV is driven back to origin by the auto-tuned controllers.}
\label{fig:exp_tuning_roll_y}
\end{figure}

This section assess the DNN-MRFT's experimental performance and persistence for synthesizing controller parameters for attitude and side-motion dynamics. Starting from a hovering state, Fig. \ref{fig:exp_tuning_roll_y} shows the two stages of the DNN-MRFT auto-tuning procedure. The identified model parameters were \(G_{att}=:\{K_{eq,att}=1.72, T_{prop}=0.0150, T_{1}=0.2005, \tau_{att} = 0.0250\}\) and \(G_{side}=:\{K_{eq,side}=4.34, T_{2}=0.3812, \tau_{side} = 0.1\}\); with the corresponding ISE optimal PD controllers \(C^{*}_{att}=:\{K_{c}=1.72, K_{d}=0.15\}\) and \(C^{*}_{side}=:\{K_{c}=1.89, K_{d}=0.56\}\). The online synthesized controllers stabilize the UAV and smoothly drive the UAV to origin point. %produce a smooth response response during hovering, and exhibit a fast settling time of \(t_{s}=\) during the step-response test. 

\label{sec:exp_results}

To evaluate the persistence of system identification, the same experiment in Fig. \ref{fig:exp_tuning_roll_y} was repeated five times. In all experiments, the DNN-MRFT approach identifies identical model parameters, which transcribes into a \(J_{cross}=0\%\) controller deterioration joint cost across the identified model parameters from all experiments. The auto-tuning experiment was also conducted with an artificial delay of 0.025 seconds added to the side-motion control loop; and the identified optimal controller parameters being \(C^{*}_{side - delay}=:\{K_{c}=1.58, K_{d}=0.56\}\). The identification results across five repetitions of the experiment with added calibrated delay were persistent with \(J_{cross}=0\%\). The controller performance deterioration cost across the two sets of experiments (with and without the added delay) was also persistent at \(J_{cross}=3\%\).

\subsection{Takeoff and Full System Identification}

Following the formulation of the take-off controllers presented in Section \ref{sec:take_off_cont}, we evaluate the capability of DNN-MRFT for full system identification and auto-tuning of UAV's starting from a landing state with no prior knowledge of system dynamics. The take-off and auto-tuning procedure is carried out in two subsequent stages: inner-loops and outer-loops identification stages. In the first stage, DNN-MRFT with the take-off controller designed in Section \ref{sec:take_off_cont} is applied to the altitude loop, corresponds to \(S\) at position \(1^*\) in Fig. \ref{fig:identification_scheme}, while placing \(S\) at position \(1\) for all other loops. Once the condition \(C\) in Eq. \eqref{eq_takeoff_algorithm} is met switch \(S\) switches to position \(1\) for altitude case as well. Once steady-state oscillations are detected on each inner-loop, a DNN identifies optimal controller parameters that are instantaneously applied to the corresponding control loop. Once inner-loop identification is complete, the switch \(S\) switches to position \(2\) such that DNN-MRFT is carried out in the same manner for the identification of side-motion loops. Similarly, once steady-state oscillations are detected on outer-loops, the corresponding DNN network identifies the dynamics and applies optimal controller by switching \(S\) to position \(3\).

\renewcommand{\arraystretch}{1.5}
\begin{table}[]
    \caption{Full system identification and tuning results on the QDrone.}
    \centering
    \begin{tabular}{|m{0.15\linewidth}|>{\centering\arraybackslash}p{0.15\linewidth}|>{\centering\arraybackslash}p{0.15\linewidth}|>{\centering\arraybackslash}p{0.15\linewidth}|>{\centering\arraybackslash}p{0.15\linewidth}|}
        \hline
        \textbf{Control Loop} & \multicolumn{2}{>{\centering\arraybackslash}m{0.3\linewidth}|}{\textbf{Full System Identification}} & \multicolumn{2}{>{\centering\arraybackslash}m{0.3\linewidth}|}{\textbf{Single Loop Identification}}\\ \cline{2-5}
            & \textbf{\(K_c\)} & \textbf{\(K_d\)} & \textbf{\(K_c\)} & \textbf{\(K_d\)} \\
        \hline
        Roll & 1.63 &  0.14 & 1.72 & 0.15 \\
        \hline
        Pitch & 1.16 &  0.13 & 1.36 & 0.13 \\
        \hline
        Altitude & 62.92 & 9.63 & 65.05 & 10.49 \\
        \hline
        x & 1.69 & 0.50 & 1.85 & 0.55 \\
        \hline
        y & 1.79 & 0.53 & 1.89 & 0.56 \\
        \hline
    \end{tabular}
    \label{tab:all_controllers}
\end{table}

Fig. \ref{fig:exp_tuning_with_90_deg}-(a, b) shows the profile of all control loops during the full-system auto-tuning experiment. The take-off controller successfully lifts the UAV while the MRFT controller stabilizes all control-loops. Once the identification phase is complete, the synthesized controllers smoothly drive the UAV back to the origin point and hold it at hover. The resulting controller parameters for the full system auto-tuning experiment are presented in Table \ref{tab:all_controllers} along with those obtained from single loop auto-tuning experiments. The similarity between the two sets of controller parameters further indicates the persistence of DNN-MRFT and validates its capability for full-system auto-tuning without prior knowledge of system dynamics.

\begin{comment}
\begin{figure}[h!]
\includegraphics[width=\linewidth]{figures/full_tuning_crop.eps} 
\caption{Full system DNN-MRFT auto-tuning experiment starting from a landing state without prior knowledge of system dynamics. (a) The UAV takes off and performs MRFT on altitude and attitude control loops. Once steady-state behaviour is observed for each control-loop, DNN-MRFT identifies the optimal controller parameters which are directly applied to control the plant. (b) DNN-MRFT is performed on both side motion control loops and tunes controllers accordingly. (c) The online tuned controllers smoothly drive the UAV back to origin and hold it at hover.}
\label{fig:exp_tuning_full}
\end{figure}
\end{comment}
 
\subsection{Trajectory Tracking Performance}

\subsubsection{Figure Eight Trajectory}
% \begin{figure}[h]
% \centering
% \begin{tabular}{>{\centering\arraybackslash}m{0.05\linewidth}|>{\centering\arraybackslash}m{0.85\linewidth}}
%  \begin{turn}{90}\textbf{Time Profile}\end{turn} &
%  \includegraphics[width=0.8\linewidth, trim={0cm 0cm 0cm 0cm}]{figures/sim_exp_8shape_xy.eps} \\
%  \hline
%  \begin{turn}{90}\textbf{Top View}\end{turn} &
%  \includegraphics[width=0.8\linewidth, trim={0cm 0cm 0cm 0cm}]{figures/sim_exp_8shape_top.eps}  \\
% \end{tabular}
% \caption{Simulation and Experimental results of the DNN-MRFT auto-tuned controller on a figure-eight trajectory.}
% \label{fig:figure_8_exp}
% \end{figure}

\begin{figure}[h]
\centering
\begin{tabular}{c|c}
 \textbf{Time Profile} & \textbf{Top View} \\ \hline
 
 \includegraphics[width=0.465\linewidth]{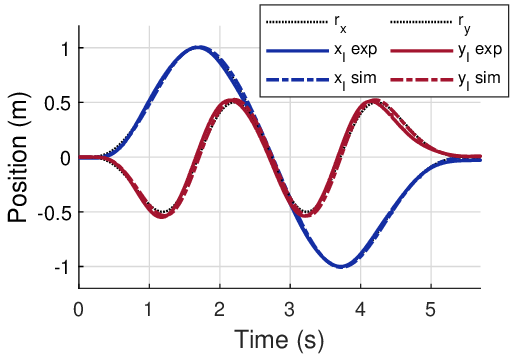} &
 \includegraphics[width=0.465\linewidth]{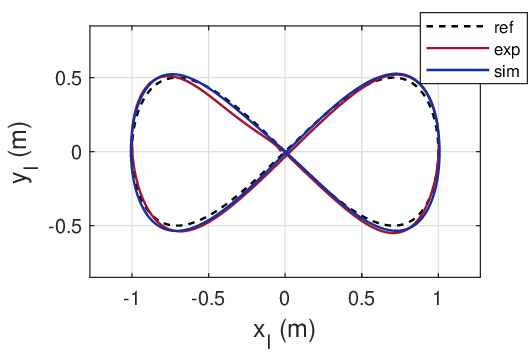}  \\
\end{tabular}
\caption{Simulation and Experimental results of the DNN-MRFT auto-tuned controller on a figure-eight trajectory.}
\label{fig:figure_8_exp}
\end{figure}
We assess the performance of the auto-tuned controller parameters shown in Table \ref{tab:all_controllers} on a figure-eight trajectory as a widely used benchmark in literature. For this purpose, we use the concept of minimum-snap trajectory optimization \cite{Mellinger2011, Richter2016} to design a figure-eight trajectory with a period of 5.5 seconds, and we provide position, velocity, and angle references for tracking. The full simulation and experimental profile of the trajectory following maneuver can be observed in Fig. \ref{fig:figure_8_exp} and a quantification of the resultant errors is provided in Table \ref{tab:exp_fig_8_compare}. The similarity between the simulation and experimental profile indicates accuracy in terms of identified model parameters, which in turn implies the optimality of the auto-tuned controllers.

\begin{comment}
\begin{figure*}[t]
\centering
\begin{tabular}{>{\centering\arraybackslash}m{0.05\textwidth}|>{\centering\arraybackslash}m{0.45\textwidth}|>{\centering\arraybackslash}m{0.45\textwidth}}
 &
 \textbf{Simulation} &
 \textbf{Experiment} \\
 \hline
 \begin{turn}{90}\textbf{time Profile}\end{turn} &
 \includegraphics[width=0.4\textwidth]{figures/sim_8shape.eps} &
 \includegraphics[width=0.4\textwidth]{figures/exp_8shape.eps} \\
 \hline
 \begin{turn}{90}\textbf{Top View}\end{turn} &
 \includegraphics[width=0.4\textwidth]{figures/sim_8_shape_top.eps} &
 \includegraphics[width=0.4\textwidth]{figures/exp_8_shape_top.eps} \\
\end{tabular}
\caption{Simulation and Experimental results of the DNN-MRFT auto-tuned controller on a figure-eight trajectory.}
\label{fig:figure_8_exp}
\end{figure*}
\end{comment}

\begin{comment}
\begin{figure*}[t]
\begin{center}
\includegraphics[width=\textwidth, trim={4cm 1.8cm 3cm 1cm}]{figures/tuning_8_shape.eps} 
\caption{Experimental results of the DNN-MRFT approach for auto-tuning UAV controllers followed by a figure-eight trajectory following. (a) Takeoff and inner-loop auto-tuning. (b) Outer-loop auto-tuning. (c) Hover. (d) Figure-eight trajectory following. }
\end{center}
\label{fig:exp_tuning_with_8shape}
\end{figure*}

\begin{figure}[h]
\includegraphics[width=\linewidth]{figures/8_shape.eps} 
\caption{Top view of figure-eight trajectory following experiment using DNN-MRFT auto-tuned controllers.}
\label{fig:exp_figure_8}
\end{figure}
\end{comment}

\renewcommand{\arraystretch}{2}
\begin{table}[t]
\caption{Experimental evaluation of tracking errors \(\norm{e(t)}\) in meters for two specifications of a figure-eight trajectory.}
    \centering
 \begin{tabular}{|p{0.006\textwidth}|p{0.19\textwidth}|>{\centering\arraybackslash}p{0.07\textwidth}|>{\centering\arraybackslash}p{0.07\textwidth}|} 
 \hline
 \multicolumn{2}{|c|}{\backslashbox{Approach}{Trajectory specs} }&  Size: \(2\times1\)m Period: 5.5s &  Size: \(4\times2\)m Period: 10.0s \\
 \hline
 \multirow{3}{4em}{\rotatebox[origin=c]{90}{Adaptive}} & Ours \((l_{mm}=0.28m)\) & \textbf{0.039} & \textbf{0.021} \\
 \cline{2-4} 
 & Ours (simulation) & 0.023 & 0.014 \\ \cline{2-4} 
 & S2R \((l_{mm}=0.13m)\) \cite{molchanov2019sim} & 0.42 & -\\
 \hline
 \multirow{2}{4em}{\rotatebox[origin=c]{90}{Manual}} & Mellinger\footnote{As reported in \cite{molchanov2019sim}} \((l_{mm}=0.13m)\) & 0.04 & - \\
 \cline{2-4} 
 & Faessler et al. \footnote{Approximated from a figure-eight trajectory with a similar maximum speed of around 3.2m/s} \cite{Faessler2018} & - & ~0.033 \\ \hline
\end{tabular}
\label{tab:exp_fig_8_compare}
\end{table}

We benchmark DNN-MRFT's performance on figure-eight trajectory tracking against two recent approaches: the S2R approach based on RL \cite{molchanov2019sim}, and differential flatness for accurate trajectory tracking approach \cite{Faessler2018}. We find the problem of S2R is closest to ours in literature, yet the nature of adaptation is quite different. We did not find other adaptive approaches applied to trajectory tracking of multirotor UAVs to compare with. We evaluate DNN-MRFT on two variants of the figure-eight trajectory. The first figure-eight trajectory we chose to perform is identical to the one reported in S2R results \cite{molchanov2019sim}, while the second trajectory is similar in size and maximum speed to the one reported in \cite{Faessler2018}. The comparison does not take into account differences in experimental setup and physical capabilities of each platform. We could not achieve speeds higher than 3.2m/s due to platform mechanical limits. Table. \ref{tab:exp_fig_8_compare} shows comparative results for the average euclidean position error \(\norm{e(t)}\) during trajectory following. The DNN-MRFT approach achieves tracking performance that is an order of magnitude better than the S2R approach \cite{molchanov2019sim}. We have also found that the tracking performance is better, or at least on par with manually tuned trajectory tracking controllers. The performance achieved by \cite{Faessler2018} relied on manual tuning and used a tailored controller to the specific trajectory being followed, which clearly requires a lot of time resources for compensation and optimization. This further confirms our conclusion that results achieved by DNN-MRFT are the new state of the art for real-time adaptive methods in trajectory tracking.

% By comparing \(\norm{e(t)}\) and \(\norm{e_{c}(t)}\), it is apparent that longitudinal errors constitute the main components of the overall position error. Longitudinal errors might arise from delays in the system response or limitation of the UAV dynamics, especially since UAV dynamics were not considered in the trajectory generation process.

\begin{figure}[h]
\begin{center}
\includegraphics[width=0.75\linewidth, trim={0cm 0cm 0cm 0cm}]{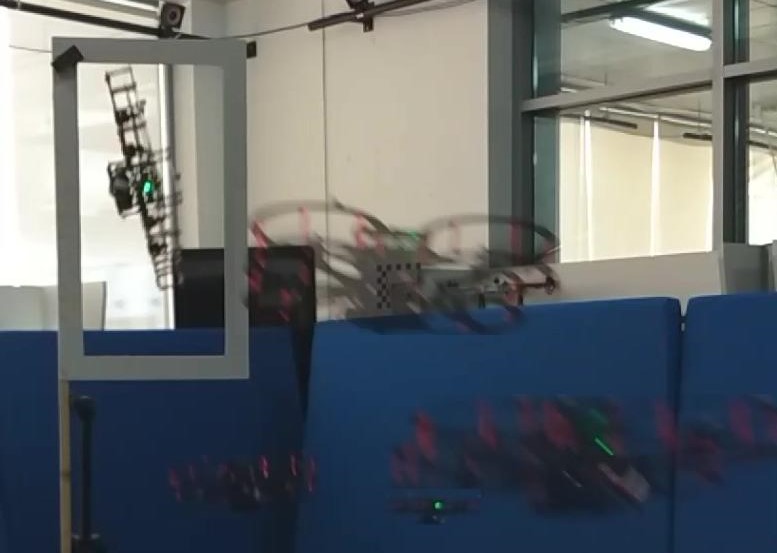}
\end{center}
\caption{The vertical window passage maneuver: the UAV passes through the window with a near \(90\deg\) roll angle.}
\label{fig:exp_90_deg}
\end{figure}

\subsubsection{Vertical Window Passage}
\begin{figure}[h]
\begin{center}
\includegraphics[width=\linewidth]{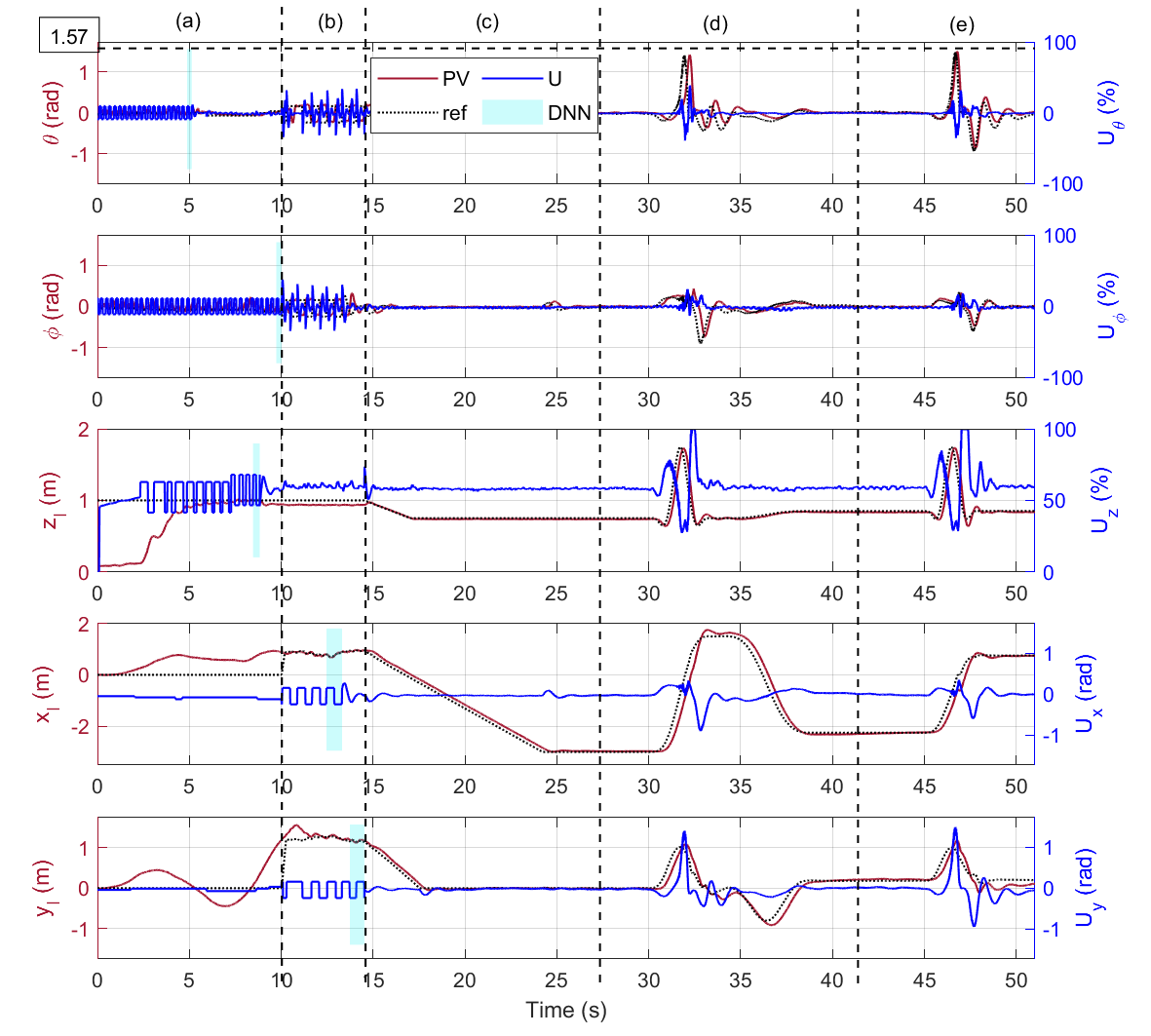}
\end{center}
\caption{Full system DNN-MRFT auto-tuning experiment starting from a landing  state without prior knowledge  of system dynamics followed by two aggressive maneuvers of vertical window passage at different speeds. (a)  The  UAV  takes  off  and  performs  MRFT  on altitude and attitude control loops. Once steady-state behaviour is  observed  for  each  control-loop,  DNN-MRFT  identifies  the optimal  controller  parameters  which  are  directly  applied  to control the plant. (b) DNN-MRFT is performed on both side motion control loops and tunes controllers accordingly. (c) The online tuned controllers smoothly drive the UAV back to origin and hold it at hover. (d) Vertical Window maneuver at 2.75 m/s. (e) Vertical Window maneuver at 1.75 m/s. A video of the full experiments can be seen in \cite{paper_video}. }
\label{fig:exp_tuning_with_90_deg}
\end{figure}

The performance of the DNN-MRFT synthesized controllers have been evaluated for an aggressive vertical-window passage maneuver as the one shown in Fig. \ref{fig:exp_90_deg}. We designed a minimum snap trajectory to pass through a vertically aligned window with a roll angle of \(90\degree\). The attitude constraint was implemented by enforcing a specific relative value between the \(x,y,\text{and } z\) acceleration components of the designed trajectory. The target window is of size \(0.3 \times 0.6 m\), leaving only 0.1 m of vertical clearance and 0.075m of horizontal clearance for the QDrone. Given this clearance and the geometry of QDrone, the minimum possible window passage velocity is 1.45 m/s. This constraint arises from the under-actuated nature of multi-rotor UAVs; which limits the attainable vertical thrust to near-zero at high pitching or rolling angles leaving the UAV at a state of free-falling. We designed two different trajectories to pass through the window at different speeds of 2.75 m/s and 1.75 m/s, with the second nearing the geometrical limit of feasible trajectories.

\begin{comment}
\begin{figure*}[t]
\includegraphics[width=\textwidth]{figures/corner_view_edit.jpg} 
\caption{Vertical window passage with the auto-tuned controllers.}
\label{fig:exp_90_deg}
\end{figure*}
\end{comment}

Fig. \ref{fig:exp_tuning_with_90_deg} shows the full state profile during both vertical window passage trajectories proceeded by the DNN-MRFT takeoff and auto-tuning phase. The auto-tuned controllers successfully maneuver the UAV and achieve an almost \(90\deg\) rolling angle at the window location. The capability of the auto-tuned controllers to execute this maneuver despite the rigid clearance constraints and the low speeds indicates that these controllers are indeed near-optimal, and constitute state-of-the-art in terms of real-time multirotor UAV auto-tuning. To the best of our knowledge, DNN-MRFT is the first UAV full auto-tuning approach to successfully perform such aggressive maneuvers without any prior knowledge of system dynamics. A video demonstration of the DNN-MRFT auto-tuning capability for the three different UAV designs listed in Table \ref{tab:uav_specs}, and vertical narrow window passage can be found in \cite{paper_video}.

\begin{comment}
\begin{figure*}[t]
\centering
\begin{tabular}{>{\centering\arraybackslash}m{0.05\textwidth}>{\centering\arraybackslash}m{0.45\textwidth}>{\centering\arraybackslash}m{0.45\textwidth}}
 &
 \textbf{2.75 m/s} &
 \textbf{1.75 m/s} \\
 \begin{turn}{90}\textbf{Front view}\end{turn} &
 \includegraphics[width=0.4\textwidth]{figures/front_view_fast_crop.jpg} &
 \includegraphics[width=0.4\textwidth]{figures/front_view_slow_crop.jpg} \\
 \begin{turn}{90}\textbf{Side view}\end{turn} &
 \includegraphics[width=0.4\textwidth]{figures/side_view_fast_crop.jpg} &
 \includegraphics[width=0.4\textwidth]{figures/side_view_slow_crop.jpg} \\
\end{tabular}
\caption{Passing through a vertical window at different speeds with the auto-tuned controllers.}
\label{fig:exp_90_deg_diff_speed}
\end{figure*}
\end{comment}

\section{Conclusion}
This paper extended the capabilities of a novel real-time system identification approach presented in \cite{ayyad2020real} referred to as DNN-MRFT. DNN-MRFT uses oscillations excited by MRFT at a specific phase, called the distinguishing phase, to identify dynamical model parameters using a DNN classifier. DNN-MRFT was extended for higher order dynamical systems in a cascaded manner. Such extension was used to identify linearized system dynamics of side translational movement having a relative degree of five with a time delay in real-time. Accuracy of system gain identification was improved by using exact solutions of Lure's systems from simulation data. DNN-MRFT precision and accuracy were validated both in simulation and experimentation. A take-off controller suitable for a wide range of UAV designs demonstrated experimental real-time identification capability. The generalized take-off controller with DNN-MRFT was verified on three different UAV designs. As a result of identification, aggressive maneuvers were possible without any form of hand tuning or biasing. The results were benchmarked against state-of-the-art and showed outstanding performance as a result of using DNN-MRFT.

For future work, we aim to evaluate DNN-MRFT for systems with other dynamical properties. We see that DNN-MRFT can be extended to MIMO systems with strong cross-loop couplings. Also, DNN-MRFT might be successful when applied to dynamical systems with large time delays. We are also aiming to handle systems with varying dynamics, where DNN-MRFT can be re-run to adapt to the change in dynamics. In such case, the identification test can be re-run to either a single or multiple control loops at which the change in dynamics occur. The time such adaptation would consume is dependant on the number of control loops to re-tune and the overall dynamics of the UAV, but should not exceed the time consumed by the full take-off and identification tests, which was limited to 15 seconds in our experimental setup. Such extensions require better understanding of the theoretical basis of information embedded on dynamical systems.

% if have a single appendix:
%\appendix[Proof of the Zonklar Equations]
% or
%\appendix  % for no appendix heading
% do not use \section anymore after \appendix, only \section*
% is possibly needed

% use appendices with more than one appendix
% then use \section to start each appendix
% you must declare a \section before using any
% \subsection or using \label (\appendices by itself
% starts a section numbered zero.)
%

% use section* for acknowledgment
\section*{Acknowledgment}
We would like to thank Quanser team for their generous and timely support. We also thank Eng. Yehya Farhoud for his help in preparing the experimental setup.

% Can use something like this to put references on a page
% by themselves when using endfloat and the captionsoff option.
\ifCLASSOPTIONcaptionsoff
  \newpage
\fi

% trigger a \newpage just before the given reference
% number - used to balance the columns on the last page
% adjust value as needed - may need to be readjusted if
% the document is modified later
%\IEEEtriggeratref{8}
% The "triggered" command can be changed if desired:
%\IEEEtriggercmd{\enlargethispage{-5in}}

% references section

% can use a bibliography generated by BibTeX as a .bbl file
% BibTeX documentation can be easily obtained at:
% http://mirror.ctan.org/biblio/bibtex/contrib/doc/
% The IEEEtran BibTeX style support page is at:
% http://www.michaelshell.org/tex/ieeetran/bibtex/
%\bibliographystyle{IEEEtran}
% argument is your BibTeX string definitions and bibliography database(s)
%\bibliography{IEEEabrv,../bib/paper}
%
% <OR> manually copy in the resultant .bbl file
% set second argument of \begin to the number of references
% (used to reserve space for the reference number labels box)
\bibliographystyle{IEEEtran}

\bibliography{bibilography.bib}

% biography section
% 
% If you have an EPS/PDF photo (graphicx package needed) extra braces are
% needed around the contents of the optional argument to biography to prevent
% the LaTeX parser from getting confused when it sees the complicated
% \includegraphics command within an optional argument. (You could create
% your own custom macro containing the \includegraphics command to make things
% simpler here.)
%\begin{IEEEbiography}[{\includegraphics[width=1in,height=1.25in,clip,keepaspectratio]{mshell}}]{Michael Shell}
% or if you just want to reserve a space for a photo:

%\appendices
%\section{Results on DJI}
%\label{app:dji_results}

\onecolumn
\appendix[Take-off Controller Performance in Simulation and Experimentation]

\begin{figure*}[h]
\centering
\includegraphics[width=0.8\textwidth,trim={0cm 1cm 0cm 0.5cm}]{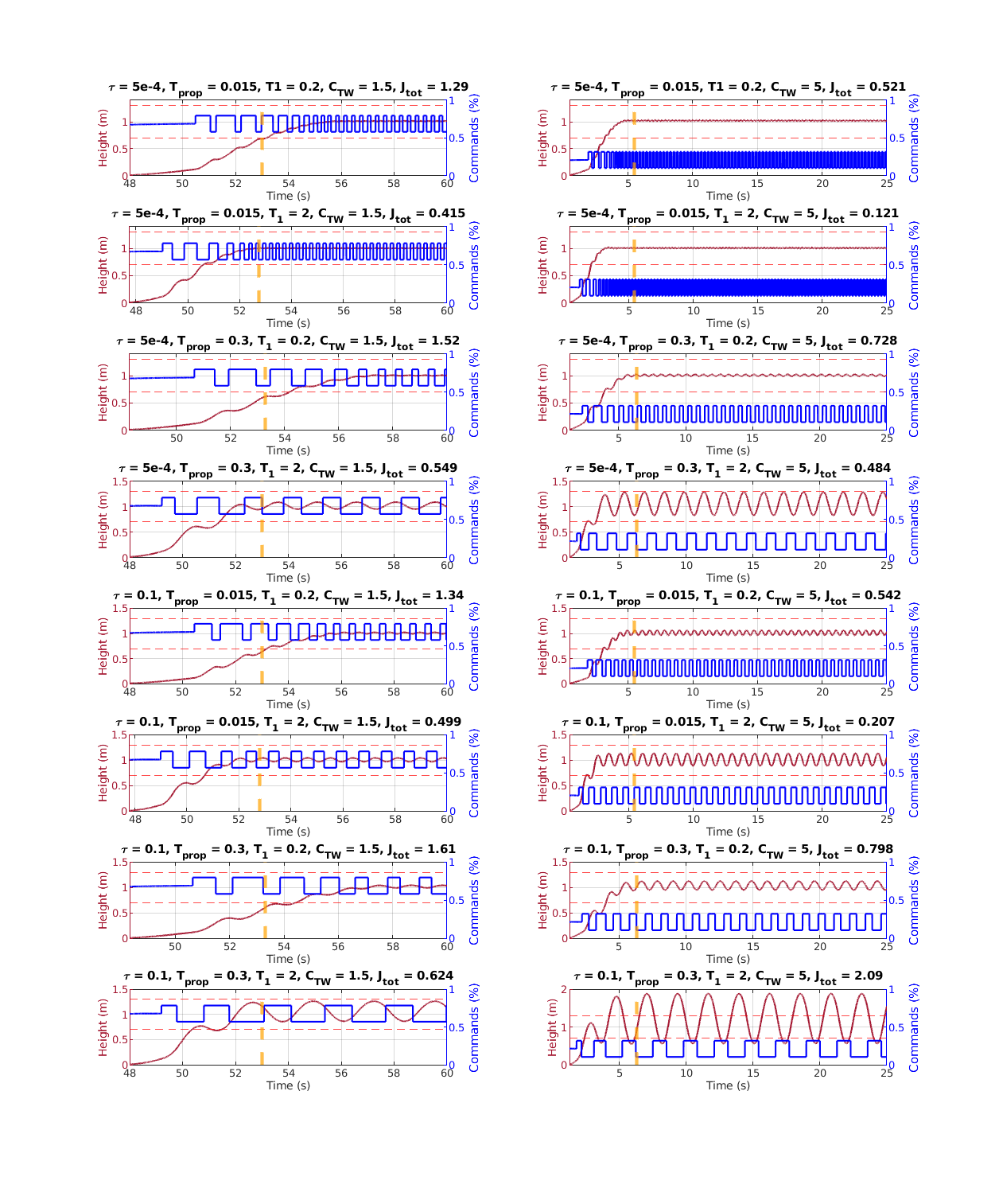} 
\caption{Response of processes at the vertices of \(D_{alt}\) for the takeoff controller with parameters in Eq. \eqref{eq_opt_takeoff_para}. The parameters at the top of each graph represents the vector \([\tau,T_{prop},T_1,C_{TW},J_{tot}]\). Vertical dashed line shows \(t_{r0}\) and the horizontal ones show \(a_{r0}\).}
\label{fig:takeoff_opt}
\end{figure*}

% you can choose not to have a title for an appendix
% if you want by leaving the argument blank
%\section{}
%Appendix two text goes here.
\begin{figure*}[h]
\centering
\includegraphics[width=0.9\textwidth,trim={0cm 0.5cm 0cm 2cm}]{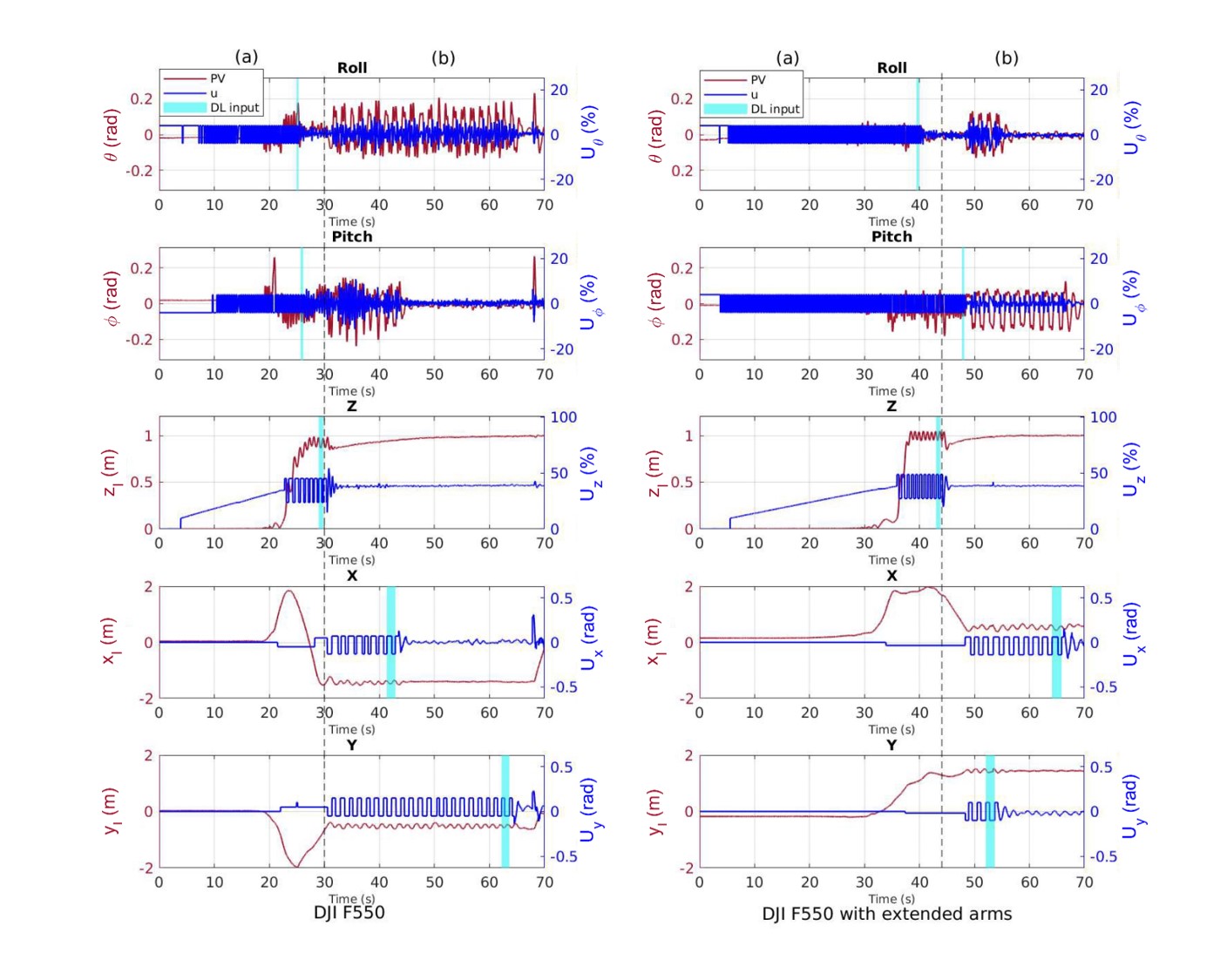} 
\caption{Full system DNN-MRFT auto-tuning experiment starting from a landing state without prior knowledge of system dynamics applied to DJI F550 custom hexarotor UAV (left column) and DJI F550 custom hexarotor UAV with extended arms (right column). In period (a) identification of inner loops parameters was performed and in period (b) identification was performed on outer loop parameters. After auto-tuning, the multirotor UAVs followed a trajectory resembling a square. Both auto-tuning experiments are shown in the video in \cite{paper_video}. }
\label{fig:exp_tuning_full_djif550}
\end{figure*}

\twocolumn

\begin{IEEEbiography}[{\includegraphics[width=1in,height=1.25in,clip,keepaspectratio]{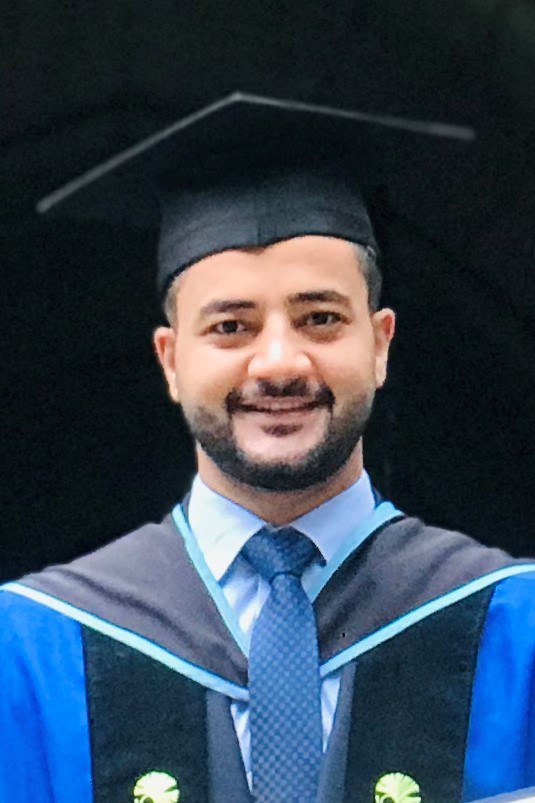}}]{Abdulla Ayyad} received his MSc. in Electrical Engineering from The University of Tokyo in 2019 where he conducted research in the Spacecraft Control and Robotics laboratory. He is currently a Research Associate in Khalifa University Center for Autonomous Robotic Systems (KUCARS) and the Aerospace Research and Innovation Center (ARIC) working on several robot autonomy projects. His current research targets the application of AI in the fields of perception, navigation, and control.
\end{IEEEbiography}

\begin{IEEEbiography}[{\includegraphics[width=1in,height=1.25in,clip,keepaspectratio]{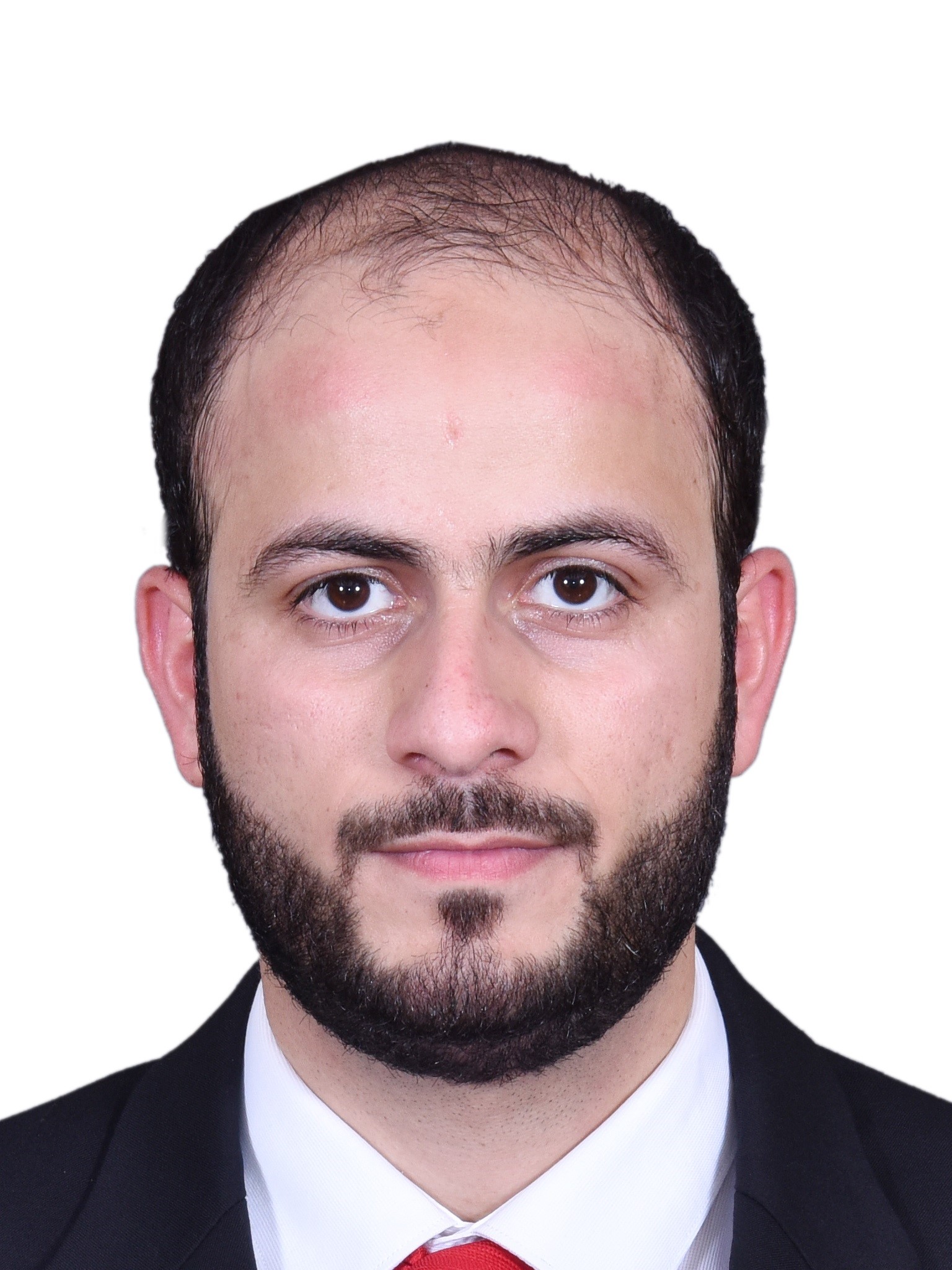}}]{Mohamad Chehadeh} received his MSc. in Electrical Engineering from Khalifa University, Abu Dhabi, UAE, in 2017. He is currently with Khalifa University Center for Autonomous Robotic Systems (KUCARS). His research interest is mainly focused on identification, perception, and control of complex dynamical systems utilizing the recent advancements in the field of AI.
\end{IEEEbiography}

\begin{IEEEbiography}[{\includegraphics[width=1in,height=1.25in,clip,keepaspectratio]{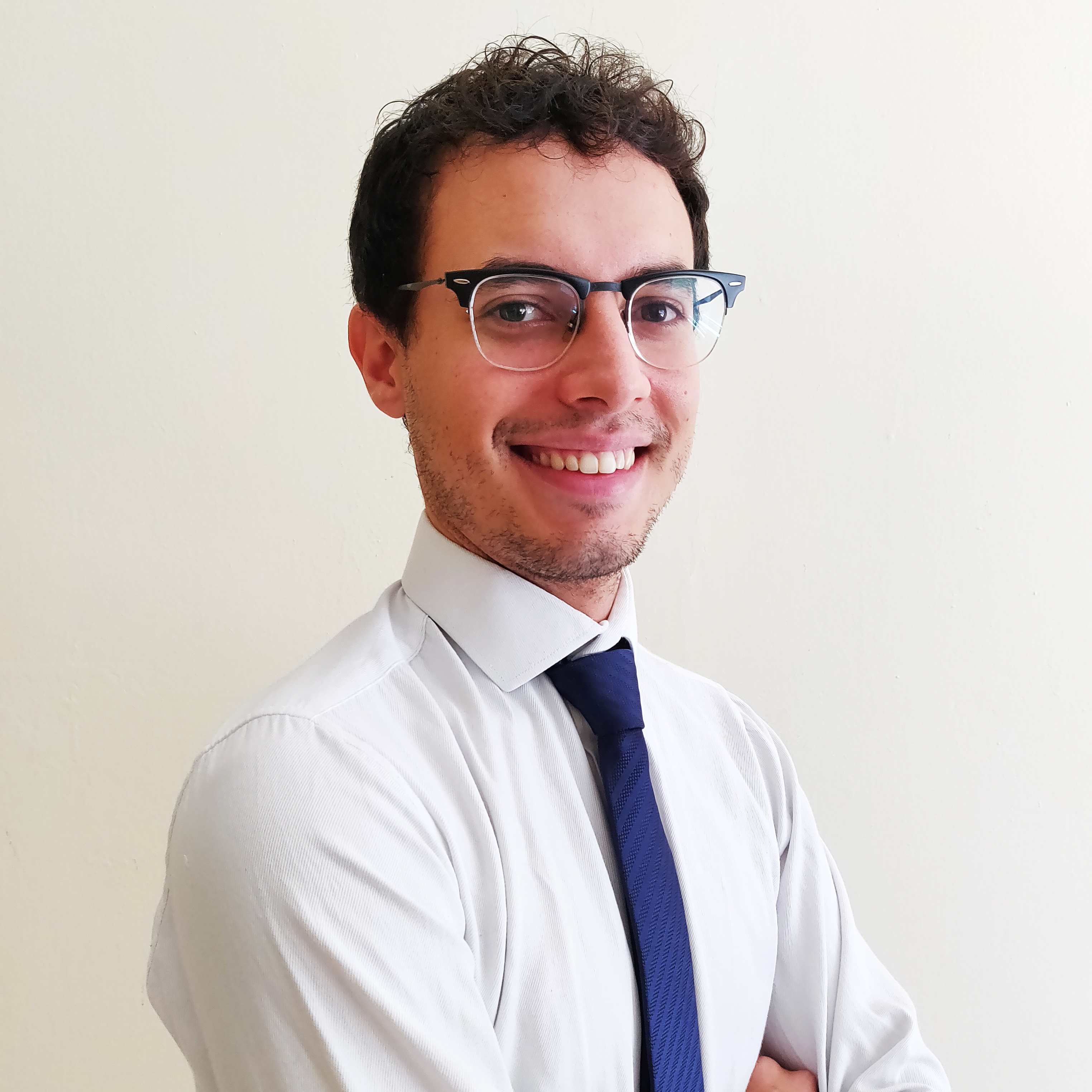}}]{Pedro Henrique Silva} received the B.Sc. degree in control engineering from the Federal University of Itajubá - Brazil, in 2017, and is currently pursuing his M.Sc. in AI from the University of Bath.
He started in the field of UAVs first as a participant on international competitions and later professionally in the agriculture industry.
He is currently a Research Associate with the Khalifa University Center for Autonomous Robotic Systems (KUCARS). His research interests include perception, control and AI applied to autonomous vehicles.
\end{IEEEbiography}

% if you will not have a photo at all:
\begin{IEEEbiography}[{\includegraphics[width=1in,height=1.25in,clip,keepaspectratio]{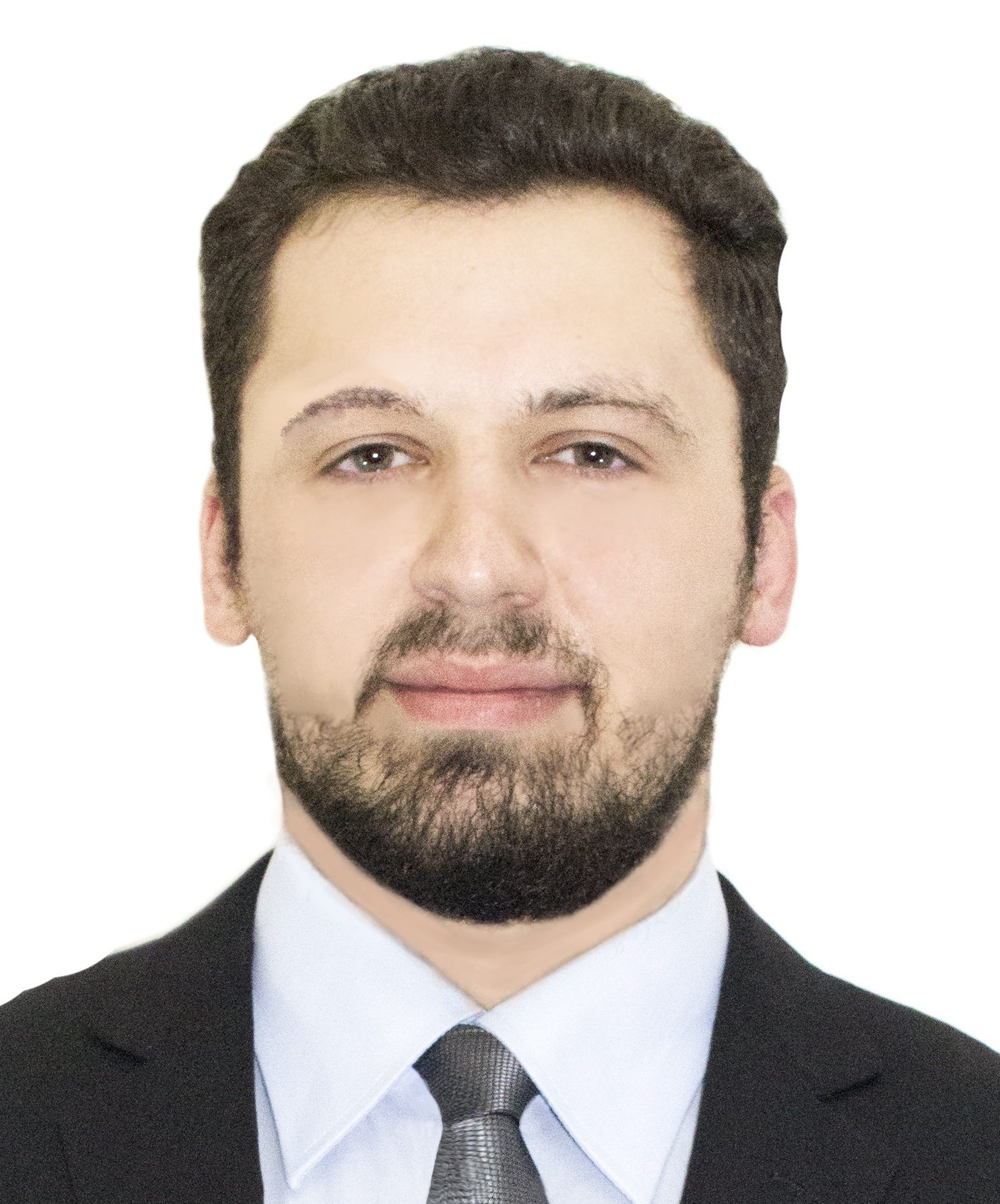}}]
{Mohamad Wahbah} received his BSc. and MSc. in Electrical Engineering from Khalifa University, Abu Dhabi, UAE. He is currently working as a research associate at Khalifa University Center for Autonomous Robotic Systems (KUCARS). Currently he is involved in perception and estimation solutions for autonomous robots.
\end{IEEEbiography}

% insert where needed to balance the two columns on the last page with
% biographies
%\newpage

\begin{IEEEbiographynophoto}{Oussama Abdul Hay}
 received his MSc. In Thermal Power and Fluids Engineering from The University of Manchester, UK. He is currently working as a research associate at Khalifa University Center for Autonomous Robotic Systems (KUCARS). His research interests are in the fields of perception and visual servoing for autonomous robots.
\end{IEEEbiographynophoto}

\begin{IEEEbiography}[{\includegraphics[width=1in,height=1.25in,clip,keepaspectratio]{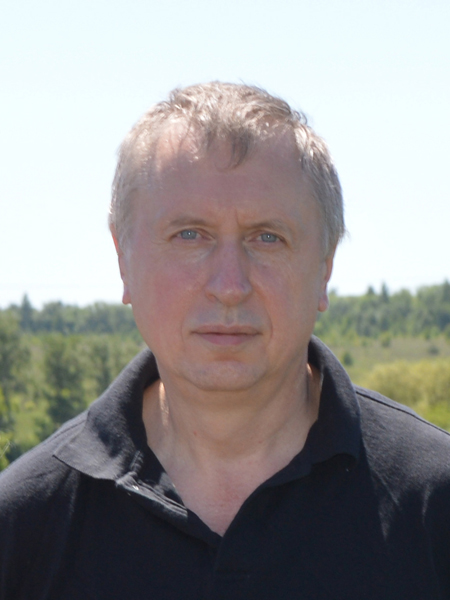}}]{Igor Boiko}
received his MSc, PhD and DSc degrees from Tula State University and Higher Attestation Commission, Russia. His research interests include frequency-domain methods of analysis and design of nonlinear systems, discontinuous and sliding mode control systems, PID control, process control theory and applications. Currently he is a Professor with Khalifa University, Abu Dhabi, UAE.
\end{IEEEbiography}

\begin{IEEEbiography}[{\includegraphics[width=1in,height=1.25in,clip,keepaspectratio]{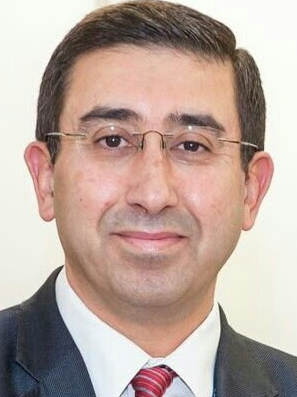}}]{Yahya Zweiri}
received the Ph.D. degree from the King’s College London in 2003. He is currently the School Director of the Research and Enterprise, Kingston University London, U.K. He is also an Associate Professor with the Department of Aerospace, Khalifa University, United Arab Emirates. He was involved in defense and security research projects in the last 20 years at the Defence Science and Technology Laboratory, King’s College London, and the King Abdullah II Design and Development Bureau, Jordan. His central research focus is interaction dynamics between unmanned systems and unknown environments by means of deep learning, machine intelligence, constrained optimization, and advanced control. He has published over 100 refereed journal and conference papers and filed six patents in USA and U.K. in unmanned systems field.
\end{IEEEbiography}

\end{document}